\documentclass{JHEP3}
\usepackage{amsmath}
\usepackage{amsfonts}
\usepackage{amssymb}

\def\bfpi{\mbox{\boldmath $\pi$}}
\newcommand{\be}{\begin{equation}}
\newcommand{\ee}{\end{equation}}

\newcommand{\ba}{\begin{eqnarray}}
\newcommand{\ea}{\end{eqnarray}}

\def\x{\;\!}

\title{Propagation of photons and massive vector mesons between a parity breaking medium and vacuum}
\author{
A. A. Andrianov$^{ab}$,  S. S. Kolevatov$^{a}$, R. Soldati$^{c}$\\
$^{a}$ V.A. Fock Department of Theoretical Physics, Sankt-Petersburg State University,\\
\phantom{$^{a}$} ul. Ulianovskaya, 198504 St. Petersburg, Russia\\
$^{b}$ High Energy Physics Group, Dept. Estructura i Constituents
de la Mat\`eria and \\\phantom{$^{a}$} Institut de Ci\`encies del Cosmos, Universitat de Barcelona,\\ \phantom{$^{a}$} Diagonal 647, 08028
Barcelona, Spain \\
$^{c}$ Dipartimento di Fisica, Universit\'a di Bologna,\\\phantom{$^{a}$}
Istituto Nazionale di Fisica Nucleare, Sezione di Bologna,\\\phantom{$^{a}$}
via Irnerio 46, 40126 Bologna, Italia\\
E-mail: \email{andrianov@icc.ub.edu, kss2005@list.ru  , roberto.soldati@infn.it}\\}
\preprint{ICCUB-11-166}

\abstract{The problem of propagation of photons and massive vector mesons in the presence of Lorenz and CPT invariance violating medium is studied  when the  parity-odd medium is bounded by a  hyperplane separating it from the vacuum. The solutions in both half-spaces are carefully discussed and in the case of space-like boundary stitched on the boundary with help of the Bogolubov transformations provided by the space-like Chern-Simons vector. The presence of two different Fock vacua is shown and the probability amplitude for transmission of particles from vacuum to parity breaking medium is calculated. We have also found classical solutions and showed that the results are consistent with ones obtained by canonical quantization formalism. In the cases, both of entrance to and of escaping from parity-odd medium, the probabilities for reflecting and passing through were found for each polarization using  the classical solutions. Finally, the propagator for each polarization is obtained in the momentum space. Boundary effects under consideration are of certain importance for registration of local parity violation in the finite volume of heavy ion fireball and/or of a star with cold axion condensate.}
\begin{document}


\section{Introduction: possible physics of time- or space-dependent pseudoscalar condensate}
The limits of validity of fundamental laws in macroscopic space-time Physics have been attracting more interest following remarkable experimental improvements,
both in laboratory research and in astrophysics \cite{Shore:2004sh}--\cite{Shao:2010wk}. Specifically in Quantum Electrodynamics the interest to possible Lorentz and CPT Invariance Violation (LIV for short) was raised up after the seminal paper \cite{Carroll:1989vb} where the very possibility to have a constant vector background generating Lorentz and CPT parity breaking in the large scale universe was  conjectured and falsified. The latter was employed to modify QED supplementing it with the Chern-Simons (CS) parity-odd lagrangian spanned on a constant CS vector. Different theoretical ways for derivation of the Carroll-Field-Jackiw Electrodynamics from the fermion matter interaction to a constant axial vector background (condensate of axial vector field, axion condensate or a gravity torsion) were considered and discussed in \cite{Jackiw:1999qq}--\cite{Casana:2008nw}.
 Later the various aspects of its signatures were discussed \cite{Andrianov:1994qv}--
\cite{Kharlanov:2009pv} although this sort of LIV has not been yet detected \cite{Kostelecky:2002hh}--\cite{Cheng:2006us}.
On the other hand, spontaneous Lorentz symmetry breaking
may occur after condensation of massless axion-like fields \cite{Andrianov:1994qv},\cite{Kostelecky:2002hh}--\cite{Alfaro:2009iv} at large space scales comparable with star and galaxies sizes.

Thus the failure in detecting a tiny violation of  Lorentz invariance and parity in the large scale universe does not exclude such effects at the level of galaxies and stars. For instance,
cold relic axions resulting from vacuum misalignment\cite{sikivie,raffelt}
in the early universe
is a viable candidate to dark matter. If we assume that
cold axions are the only contributors to the matter density of the universe
apart from ordinary baryonic matter its density must be \cite{wmap,wmap1,wmap2}
$
\rho\simeq 10^{-30}{\rm g} {\rm cm}^{-3}\simeq 10^{-46} {\rm GeV}^4.
$
Of course dark matter is not uniformly distributed, its distribution
presumably follows that of visible matter. One may think also of an axion background accumulated by very dense stars like neutron ones or even of bosonic axion stars \cite{mielke}. Slowly varying axion background or condensed axions can be in principle discovered as such a background induces the high-energy photon decays into dilepton pairs \cite{axion1} and, in turn,
photon emission by charged particles \cite{axion2,Espriu:2011vj,Espriu:2010bj,Espriu:2010ip}.

Another interesting area for observation of parity breaking is the heavy ion physics. Recently several experiments in heavy ion collisions  have indicated an abnormal yield of lepton pairs of
invariant mass $< 1$ GeV in the region of small rapidities and moderate transversal momenta \cite{ceres,ceres1,NA60,phenix,HADES} (see, the reviews \cite{lapidus,tserruya}).
This phenomenon is stronger for central collisions with moderate transversal momenta of dileptons. It happens both for $e^+e^-$ and dimuon pairs  and a possible explanation of this enhancement is outlined in \cite{aaep}. It was conjectured that the effect may be a manifestation of
local parity breaking (LPB) in colliding nuclei due to generation of pseudoscalar, isosinglet or neutral isotriplet, classical
background whose magnitude depends on the dynamics of the collision. Theoretical reasons to generate an isotriplet pseudoscalar condensate at large baryon densities have been given in \cite{anesp,anesp1}. In addition to, in \cite{kharzeev,zhit,kharzeev1,kharzeev2} it has been suggested that for peripheral interactions  a complementary phenomenon of the so-called Chiral Magnetic Effect
 should occur. It is triggered by  an isosinglet pseudoscalar background
as the result of large-scale fluctuation of topological charge and has been studied by lattice QCD simulations \cite{lattice} and nearly detected in the STAR experiments on RHIC \cite{star,star2}.

In the occurrences of axion-like background in  astrophysics or  heavy ion physics the existence of a boundary between the parity-odd medium and the vacuum is quite essential. For star condensed axions there is evidently a boundary where axion background disappears and photons distorted by it escape to vacuum.
In the presence of such a space boundary  not all the photons penetrate it and partially a reflection arises which will be described in details in Sec.\ref{refl}. As well in heavy ion collisions the outcome of the photon/vector meson decays, say, into lepton pairs generated by slowly decreasing pseudoscalar background inside of the fireball can be normally registered in vacuum or after freeze-out when dilepton pairs are outside of the parity breaking medium.

Thus the examination of how an axion/pion background in a bounded volume can influence on photon and massive vector mesons propagating through a boundary to vacuum represents a real interest for detecting parity odd properties of a medium. In this work we analyze a simplified model with two semi-infinite spaces,  one of which is filled by a pseudoscalar matter and another one to be a normal vacuum. Such a model may shed light on the processes of photon and vector meson emission in the crust of axion reach neutron stars or on the boundary of fireballs when wave lengths of vector states are much less than the size of a distorted vacuum with parity breaking background. It is defined in the next section and its spectrum depending on polarizations and canonical quantization is examined in details. In Sec.3 we show that for a purely space-arrowed CS vector such an electrodynamics can be consistently quantized having the two different vacuum states related by Bogolyubov transformation. In  Sec.4 the matching of vector boson fields between parity-odd CS medium and vacuum is elaborated in details for a space-like boundary. The transmission and reflection
of massive vector mesons when propagating outside of parity-odd medium to vacuum and in the opposite direction are calculated. In Subsec. \ref{green} the Green function for this propagation is reconstructed. Conclusions and perspectives for observation of parity odd media are summarized in the last section where also the geometry of distorted photon decays is discussed for how to serve
as a possible signature for detecting parity breaking media by the anisotropy in emission
of dilepton pairs.

\section{Vector Fields in a Pseudoscalar Background}
We start from the Lagrange density which describes the propagation of a vector
field in the presence of a  pseudoscalar axion-like background,
\begin{eqnarray}
{\mathcal L} &=&  -\,{\textstyle\frac14}\,F^{\alpha\beta}(x)F_{\alpha\beta}(x)
-\,{\textstyle\frac14}\,{g}\,F^{\mu\nu}(x)\widetilde F_{\mu\nu}(x)\,a_{c\ell}(x)/M
\nonumber\\
&+&  {\textstyle\frac12}\,m^2\,A_\nu(x)A^\nu(x) + A^\mu(x)\,\partial_\mu B(x) +
{\textstyle\frac12}\,\varkappa\,B^2(x),\label{lagrangian1}
\end{eqnarray}
where $A_\mu$ and $a_{c\ell}$ stand for the vector and
background  pseudoscalar fields respectively,
$\widetilde F^{\mu\nu}={\textstyle\frac12}\,
\varepsilon^{\,\mu\nu\rho\sigma}\,F_{\,\rho\sigma}$ is the dual field strength,
while $B$ is the auxiliary St\"uckelberg scalar field with $\varkappa\in\mathbb R$. The positive dimensionless coupling
$g>0$ and the (large) mass parameter $M\gg m$ do specify the intensity and the scale of the pseudoscalar-vector
interaction. Notice that we have included
the Proca mass term for the vector field because, as it is
discussed in \cite{aaep}, the latter is required to account for the strong interaction effects in heavy ion collisions supported by massive vector mesons ($\rho,\omega,\ldots$) in addition to
photons.
Moreover, as thoroughly debated in \cite{AACGS2010}, the mass term for the vector field
appears to be generally necessary to render the dynamics self-consistent in the presence of a
Chern-Simons lagrangian and is generally induced by radiative corrections from the fermionic matter
lagrangian\footnote{We also leave the room for the photon mass generation in a plasma like medium.}.
The auxiliary St\"uckelberg lagrangian,
which further violates gauge invariance beyond the mass term for the vector field,
has been introduced to provide -- just owing to the renowned St\"uckelberg trick --
the simultaneous occurrences of power counting renormalizability and perturbative unitarity
for a general interacting theory. Moreover, its presence allows for a smooth massless limit of the
quantized vector field.
\subsection{The Chern-Simons Limit for the Pseudoscalar Background}
We shall consider a slowly varying classical pseudoscalar background of the kind
\begin{equation}
a_{c\ell}(x)=\frac{M}{g}\,\zeta_\lambda x^{\lambda}\,\theta(-\,\zeta\cdot x)\label{background}
\end{equation}
where $\theta(\cdot)$ is the Heaviside step distribution,
in which a fixed constant four vector $\zeta^\mu$ with dimension of a mass has been introduced,
in a way to violate Lorentz and CPT invariance in the Minkowski half space
$\zeta\cdot x<0\,.$ In what follows we shall suppose that $\zeta^2\not=0\,.$
If we now insert the specific form (\ref{background}) of the pseudoscalar background
in the pseudoscalar-vector coupling lagrangian we can write
\begin{eqnarray}
-\,{\textstyle\frac14}\,F^{\mu\nu}(x)\widetilde F_{\mu\nu}(x)\,\zeta_\lambda x^{\lambda}\,\theta(-\,\zeta\cdot x)
&=& {\textstyle\frac12}\,\zeta_\mu A_\nu(x)\widetilde F^{\mu\nu}(x)\,\theta(-\,\zeta\cdot x)\nonumber\\
&-& \partial_\mu \left[\,{\textstyle\frac12}\,A_\nu(x)\widetilde F^{\mu\nu}(x)\,
\zeta_\lambda x^{\lambda}\,\theta(-\,\zeta\cdot x)\,\right]
\end{eqnarray}
The very last term in the RHS of the above equality is evidently a boundary term, its contribution to the Action
being reduced for the Gau\ss\ theorem to
\begin{displaymath}
\int_\Omega\mathrm{d}^4x\;\partial_\mu \left[\,{\textstyle\frac12}\,A_\nu(x)\widetilde F^{\mu\nu}(x)\,
\zeta_\lambda x^{\lambda}\,\theta(-\,\zeta\cdot x)\,\right]
={\textstyle\frac12}\int_{\partial\Omega}\mathrm{d}\sigma_\mu\;A_\nu(x)\widetilde F^{\mu\nu}(x)\,
\zeta_\lambda x^{\lambda}\,\theta(-\,\zeta\cdot x)
\end{displaymath}
where $\Omega$ is an arbitrary domain of the Minkowski space-time that is bounded by the \textit{initial}
and \textit{final} three dimensional
space-like oriented surfaces $\partial\Omega=\Sigma_\imath\cup\Sigma_{f}\,.$ Hence the boundary term won't
contribute to the Euler-Lagrange field equations iff
\[\left.A_\nu(x)\widetilde F^{\mu\nu}(x)\,
\zeta_\lambda x^{\lambda}\,\theta(-\,\zeta\cdot x)\right|_{\Sigma_\imath}
=\left.A_\nu(x)\widetilde F^{\mu\nu}(x)\,
\zeta_\lambda x^{\lambda}\,\theta(-\,\zeta\cdot x)\right|_{\Sigma_f}\equiv0\]
which entails a particular fall down of the vector potential and field strength
for large space-like separations in the half space-time $\zeta\cdot x<0\,.$ In such a circumstance
we can derive the field equations from the equivalent Lagrange density
\begin{eqnarray}
{\mathcal L} &=&  -\,{\textstyle\frac14}\,F^{\alpha\beta}(x)F_{\alpha\beta}(x)
+ {\textstyle\frac12}\,\zeta_\mu A_\nu(x)\widetilde F^{\mu\nu}(x)\,\theta(-\,\zeta\cdot x)
\nonumber\\
&+&  {\textstyle\frac12}\,m^2\,A_\nu(x)A^\nu(x) + A^\mu(x)\,\partial_\mu B(x) +
{\textstyle\frac12}\,\varkappa\,B^2(x)\label{lagrangian2}
\end{eqnarray}
in which the gauge invariance is badly broken by all the terms but the one, i.e. the Maxwell's radiation lagrangian.
Then the field equations read
\begin{eqnarray}
\left\lbrace
\begin{array}{cc}
\partial_\lambda F^{\,\lambda\nu}
+\ m^2\,A^\nu + \zeta_\alpha\widetilde F^{\,\alpha\nu} +
\partial^{\,\nu} B=0 & \qquad{\rm for}\ \zeta\cdot x<0\\
\partial_\lambda F^{\,\lambda\nu}
+\ m^2\,A^\nu + \partial^{\,\nu} B=0 & \qquad{\rm for}\ \zeta\cdot x>0\\
\partial_\nu A^\nu\;=\;\varkappa\,B
\end{array}\right. \label{Euler_Lagrange}
\end{eqnarray}
After contraction of the first pair of the above set of field equations with $\partial_\nu$ we find
\begin{equation}
\left(\Box+\varkappa\,m^{2}\right)B(x)=0
\label{aux}
\end{equation}
whence it follows that the auxiliary St\"uckelberg field is always a decoupled
unphysical real scalar field, which is never affected by the pseudoscalar
classical background $\forall\,\varkappa\in\mathbb R$. From now on we shall select
the simplest choice $\varkappa=1$ that leads to the Klein-Gordon equation
for the auxiliary field, together with
\begin{eqnarray}
\left\lbrace
\begin{array}{cc}
\Box A^\nu(x)  +  m^2\,A^\nu(x) = \varepsilon^{\,\nu\alpha\rho\sigma}\,\zeta_\alpha\,\partial_\rho A_\sigma(x)
& \qquad{\rm for}\ \zeta\cdot x<0 \\
\Box A^\nu(x)  +  m^2\,A^\nu(x)=0
& \qquad{\rm for}\ \zeta\cdot x>0\\
\partial_\nu A^\nu(x)=B(x)
&\qquad\quad \left(\Box+m^{2}\right)B(x)=0
\end{array}\right. \label{EL2}
\end{eqnarray}

In order to find the most general solution of the above linear equations (\ref{EL2})
we extend the equations in half-spaces to the entire plane and turn to the momentum space
\[
A^\nu(x)=\int\frac{{\rm d}^4k}{(2\pi)^{3/2}}\,{\rm \mathbf{a}}^\nu(k)\,{\rm e}^{\,-i k\cdot x}
\qquad\qquad
B(x)=\int\frac{{\rm d}^4k}{(2\pi)^{3/2}}\,{\rm b}(k)\,{\rm e}^{\,-ik\cdot x}
\]
so, the second and third line equations (\ref{EL2}) for entire space-time
\begin{equation}
\left\lbrace
\begin{array}{cc}
 \Box A^\nu(x)  +  m^2\,A^\nu(x)=0\\
 \qquad \partial_\nu A^\nu(x)=B(x)  \qquad\quad \left(\Box+m^{2}\right)B(x)=0
 \end{array}\right.
\label{MEL2}
\end{equation}
can be written in the momentum space,
 \begin{equation}
\left\lbrace
\begin{array}{cc}
\left[\,g^{\,\lambda\nu}\left(\,k^2-m^2\,\right) - k^{\,\lambda}k^{\,\nu}\,\right]
 \mathbf{a}_\lambda(k) + i\,k^{\,\nu}\,\mathrm b(k)=0\\
 \\
k^{\,\lambda}\,\mathbf{a}_\lambda(k) = i\,\mathrm b(k)
\qquad\qquad\quad \left(k^2-m^{2}\right)\mathrm b(k)=0
\end{array}\right.
\end{equation}
The general solutions of these field equations are the well known
Proca-St\"uckelberg vector and auxiliary ghost scalar quantum free fields, viz.,
\begin{eqnarray}
A^{\mu}(x) &=& A^{\mu}_{\rm PS}(x) - \partial^{\,\mu}B(x)/m^2
\\
A^{\mu}_{\rm PS}(x) &=& \int\mathrm d^3 \mathbf k\sum_{r=1}^3\
\left[\,\mathbf{a}_{\,{\bf k}\,,\,r}\,u^{\,\mu}_{\,{\bf k}\,,\,r}(x)
+ \mathbf{a}^\dagger_{\,{\bf k}\,,\,r}\,
u^{\,\mu\,\ast}_{\,{\bf k}\,,\,r}(x)\,\right]
\qquad\quad\partial_{\mu}A^{\mu}_{\rm PS}(x)=0\\
B(x)&=&m\int\mathrm d^3 \mathbf k\,\Big[\,b_{\bf k}\,u_{\bf k}(x) +
b^\dagger_{\bf k}\,u^\ast_{\bf k}(x)\,\Big]\\
u^\nu_{{\bf k}\,,\,r}(x)&=&[\,(2\pi)^3\,2\omega_{\,\bf k}\,]^{-1/2}\,
e_{\,r}^{\,\nu}({\bf k})\,
\exp\{-\,i\,\omega_{\,\bf k} x^0 + i\,{\bf k}\cdot{\bf x}\}\qquad (\,r=1,2,3\,)\\
u_{\,\bf k}(x)&=&[\,(2\pi)^3\,2\omega_{\,\bf k}\,]^{-1/2}\,
\exp\{-\,i\,\omega_{\,\bf k} x^0 + i\,{\bf k}\cdot{\bf x}\}\qquad\qquad
\omega_{\,\bf k}\ \equiv\ \sqrt{{\bf k}^{2}+m^2}
\end{eqnarray}
where the creation destruction operators fulfill the canonical commutation relations
\begin{equation}
 [\,\mathbf a_{\,{\bf k}\,,\,r}\,,\,\mathbf a^\dagger_{\,{\bf k}^\prime\,,\,s}\,]=\delta({\bf k}-{\bf k}^\prime)\,\delta_{rs}
\qquad\quad
[\,b^\dagger_{\bf k}\,,\,b_{\,{\bf k}^\prime}\,]=\delta({\bf k}-{\bf k}^\prime)
\end{equation}
all the remaining commutators being equal to zero.
The three linear polarization real vectors do satisfy the orthonormality and closure relations
on the mass shell $k^2=m^2\,:$ namely,
\begin{eqnarray}
k_{\mu}e_{\,r}^{\,\mu}({\bf k})=0\qquad
 -\,g_{\,\mu\nu}\;e_{\,r}^{\,\mu}({\bf k})\,
e_{\,s}^{\,\nu}({\bf k})=\delta_{\,rs}
\qquad
\sum_{r=1}^3\,e_{\,r}^{\,\mu}({\bf k})\,
e_{\,r}^{\,\nu}({\bf k})
= -\;g^{\;\mu\nu} + \frac{k^{\mu}k^{\nu}}{m^2}
\end{eqnarray}
while the vector plane wave functions fulfill
\begin{equation}
\partial_{\mu}u^{\,\mu}_{\,{\bf k}\,,\,r}(x)=0
\qquad\quad\forall\,\mathbf k\in\mathbb R^3\,,\,r=1,2,3
\end{equation}
\begin{equation}
\left( u^\mu_{{\bf k}\,,\,r}\,,\,u^\nu_{{\bf p}\,,\,s}\right)=
\int\mathrm d^3 \mathbf x\,u^{\mu\ast}_{{\bf k}\,,\,r}(t,\mathbf x)
\,i\overleftrightarrow{\partial_0}u^\nu_{{\bf p}\,,\,s}(t,\mathbf x)
=\delta({\bf k}-{\bf p})\,e_{\,r}^{\,\mu}({\bf k})\,e_{\,s}^{\,\nu}({\bf p})
=-\,\left( u^{\mu\ast}_{{\bf k}\,,\,r}\,,\,u^{\nu\ast}_{{\bf p}\,,\,s}\right)
\end{equation}
\begin{equation}
\left( u^{\mu\ast}_{{\bf k}\,,\,r}\,,\,u^\nu_{{\bf p}\,,\,s}\right)=
\left( u^{\mu}_{{\bf k}\,,\,r}\,,\,u^{\nu\ast}_{{\bf p}\,,\,s}\right)=0
\label{orthonorma}
\end{equation}

For the half-space $\zeta\cdot x<0$  one can perform the similar extension of Maxwell Electrodynamics involving the Chern-Simons term onto entire space-time. In the momentum space,
 \begin{equation}
\left\lbrace
\begin{array}{cc}
\left[\,g^{\,\lambda\nu}\left(\,k^2-m^2\,\right) - k^{\,\lambda}k^{\,\nu} +
i\,\varepsilon^{\,\lambda\nu\alpha\beta}\,\zeta_\alpha\,k_\beta\,\right]
 \mathbf{a}_\lambda(k) + i\,k^{\,\nu}\,\mathrm b(k)=0
 \\
k^{\,\lambda}\,\mathbf{a}_\lambda(k) = i\,\mathrm b(k)
\qquad\qquad\quad \left(k^2-m^{2}\right)\mathrm b(k)=0
\end{array}\right.
\label{MEL2}
\end{equation}

The general solutions of the equations for the Maxwell-Chern-Simons
free quantum field, have been extensively discussed and applied in
\cite{AACGS2010,AEGS2009} for the massive case and in \cite{AGS2002} for the massless case.
However, in the light of the present applications it's better to shortly overview this topic.
\subsection{Canonical Quantization of the Maxwell-Chern-Simons Field}
To be definite, let us first recall the construction of the so called chiral or birefringent polarization vectors
for the Maxwell-Chern-Simons (MCS) vector field. Here we aim to develop a rather general frame which
could allow to readily interplay among the massive, massless, temporal and spatial cases, as we shall specify in the sequel.
The starting point is the rank-two symmetric matrix
\cite{AACGS2010}
\begin{equation}
S^{\nu}_{\phantom{\nu}\lambda}\equiv\varepsilon^{\mu\nu\alpha\beta}\,\zeta_{\alpha}\,k_{\beta}\,
\varepsilon_{\mu\lambda\rho\sigma}\,\zeta^{\,\rho}k^{\sigma}=
\delta^{\,\nu}_{\;\lambda}\,{\mbox{\tt D}}
+ k^{\nu}\,k_{\lambda}\,\zeta^2 + \zeta^{\,\nu}\,\zeta_{\lambda}\,k^2
- \zeta\cdot k\,(\zeta_{\lambda}\,k^{\nu} + \zeta^{\nu}\,k_{\,\lambda})
\end{equation}
where
\[
{\mbox{\tt D}}\;\equiv\;(\zeta\cdot k)^2-\zeta^2\,k^2\;=\;\textstyle\frac12\;S^{\nu}_{\phantom{\nu}\nu}
\]
in such a manner that we find
\begin{equation}
S^{\,\nu}_{\;\lambda}\,\zeta^{\lambda}=S^{\,\nu}_{\;\lambda}\,k^{\lambda}\;=\;0\qquad\quad
S^{\,\mu\nu}\,S_{\,\nu\lambda}\;=\;{\mbox{\tt D}}\,S^{\mu}_{\phantom{\mu}\lambda}\qquad\quad
S^{\nu}_{\phantom{\nu}\nu}\;=\;2\,{\mbox{\tt D}}
\end{equation}
while
\begin{equation}
S^{\,\mu\lambda}\,\varepsilon_{\lambda\nu\alpha\beta}\,\zeta^\alpha k^{\beta}\;
=\;{\mbox{\tt D}}\;\varepsilon^{\,\mu}_{\phantom{\mu}\nu\alpha\beta}\,\zeta^\alpha k^{\beta}
\end{equation}
Notice that for a temporal Chern-Simons vector $\zeta_{\mu}=(\zeta_0,0,0,0)$
we find ${\mbox{\tt D}}=\zeta_0^2\,{\bf k}^2\ge0\,,$ while for a spatial one like e.g.
$\zeta_{\mu}=(0,-\zeta_x,0,0)$ we get ${\mbox{\tt D}}=\zeta_x^2(k_0^2-k_2^2-k_3^2)\ge0$
in the restricted causal domain $k_0^2\ge k_2^2+k_3^2$ which lies inside
\footnote{Actually it corresponds to the light cone section
$\mathfrak D\equiv\{k_0^2\mathbf k^2\}\cap\{k_1=0\}$.}
the causal cone $k^2\ge0$.
Then, to our purpose, it is convenient to introduce the
two orthonormal, one dimensional, hermitian projectors
\begin{equation}
\bfpi^{\,\mu\nu}_{\,\pm}\equiv
\frac{S^{\,\mu\nu}}{2\,{\mbox{\tt D}}}\;
\pm\;\frac{i}{2}\,
\varepsilon^{\mu\nu\alpha\beta}\,\zeta_{\alpha}\,k_{\beta}\,{\mbox{\tt D}}^{\,-\frac12}
=\left(\bfpi^{\,\nu\mu}_{\,\pm}\right)^\ast=\left(\bfpi^{\,\mu\nu}_{\,\mp}\right)^\ast
\qquad\quad(\mbox{\tt D}>0)
\end{equation}
It is worthwhile to observe that the addendum involving the Levi-Civita symbol is always imaginary for a temporal
Chern-Simons vector $\zeta_{\mu}=(\zeta_0,0,0,0)$, while for the  spatial one $\zeta_{\mu}=(0,-\zeta_x,0,0)$ it is imaginary
in the domain $k_0^2\ge k_2^2+k_3^2$.
The above pair of chiral projectors actually encodes the occurrence of birefringence or vacuum Faraday's effect
and enjoys the following useful properties $\forall\,k^{\,\mu}=(k_0,{\bf k})$: namely,
\begin{eqnarray}
\bfpi^{\,\mu\nu}_{\,\pm}\;\zeta_{\nu}=\bfpi^{\,\mu\nu}_{\,\pm}\;k_{\nu}=0
\qquad\quad g_{\,\mu\nu}\,\bfpi^{\,\mu\nu}_{\,\pm}\;=\;1
\label{projectors1}\\
\bfpi^{\,\mu\lambda}_{\,\pm}\,\bfpi_{\,\pm\,\lambda\nu}\ =\
\bfpi^{\,\mu}_{\,\pm\,\nu}\qquad\quad
\bfpi^{\,\mu\lambda}_{\,\pm}\,\bfpi_{\,\mp\,\lambda\nu}\ =\ 0
\label{projectors2}
\end{eqnarray}
\[
\bfpi^{\,\mu\nu}_{\,+}\;+\;\bfpi^{\,\mu\nu}_{\,-} = S^{\mu\nu}/\,{\mbox{\tt D}}
\]
\[
\bfpi^{\,\mu\nu}_{\,+}\;-\;\bfpi^{\,\mu\nu}_{\,-}=
i\varepsilon^{\mu\nu\alpha\beta}\,\zeta_{\alpha}k_{\beta}\,{\mbox{\tt D}}^{\,-\frac12}
\]
A couple of chiral polarization vectors for the Maxwell-Chern-Simons free vector field can be constructed out of some
tetrad of constant quantities $\epsilon_\nu$, taking into account that we have
\begin{eqnarray}
 \bfpi^{\,\mu\lambda}_{\,\pm}\,\epsilon_\mu\epsilon_\lambda=
{\mbox{\tt D}}\,\epsilon^2+\zeta^2(\epsilon\cdot k)^2=
[\,(\zeta\cdot k)^2-\zeta^2\,k^2\,]\,\epsilon^2+\zeta^2(\epsilon\cdot k)^2
\end{eqnarray}
For example, if we choose $\epsilon^{\nu}=(0,0,0,1)$, for the spatial Chern-Simons vector $\zeta_{\mu}=(0,-\,\zeta_x,0,0)$
we find
\begin{displaymath}
 \bfpi^{\,\mu\lambda}_{\,\pm}\,\epsilon_\mu\epsilon_\lambda=
\frac{k_2^2-k_0^2}{k_0^2 - k_2^2 - k_3^2}= -\,1 + \frac{k_3^2}{k_3^2 + k_2^2 - k_0^2}<0
\qquad\forall\,k^{\,\mu}=(k_0,\mathbf k),\quad k_0^2 \ge k_2^2 - k_3^2 .
\end{displaymath}

Alternatively, had we opted for the symmetric choice
$\bar\epsilon^{\,\nu}=(0,1,1,1)/\sqrt3$, then for the temporal Chern-Simons vector
$\zeta_{\mu}=(\zeta_0,0,0,0)$ we get
\begin{displaymath}
 \bfpi^{\,\mu\lambda}_{\,\pm}\,\bar\epsilon_\mu\bar\epsilon_\lambda=
-\,\frac12 + \frac{(k_1+k_2+k_3)^2}{6\mathbf k^2}<0
\qquad\forall\,k^{\,\mu}=(k_0,\mathbf k)
\end{displaymath}
Hence in both cases we can always build up a pair of space-like, complex, birefringent, chiral polarization vectors
\begin{equation}
\varepsilon_{\,\pm}^{\,\mu}(k)=
\left\lbrace
 \begin{array}{cc}
\bfpi^{\,\mu\lambda}_{\,\pm}\,\epsilon_\lambda\,\left[(k_2^2 - k_0^2)/(k_3^2 + k_2^2 - k_0^2)\right]^{-\frac12}
&  {\rm for}\quad\zeta_{\mu}=(0,-\,\zeta_x,0,0)\\
\bfpi^{\,\mu\lambda}_{\,\pm}\,\bar\epsilon_\lambda\,\left[\frac12-(k_1+k_2+k_3)^2/6\mathbf k^2\,\right]^{-\frac12}
&  {\rm for}\quad\zeta_{\mu}=(\zeta_0,0,0,0)
 \end{array}\right.
\end{equation}
For $\mbox{\tt D}>0$ this couple of chiral polarization vectors satisfy the conjugation and orthonormality relations
\begin{displaymath}
\varepsilon_{\,\pm}^{\,\mu\ast}(k)=\varepsilon_{\,\mp}^{\,\mu}(k)\qquad\quad
-\,g_{\mu\nu}\;\varepsilon^{\,\mu\ast}_{\pm}(k)\,\varepsilon^{\,\nu}_{\pm}(k)
=1\qquad\quad
g_{\mu\nu}\,\varepsilon^{\,\mu\ast}_{\pm}(k)\,\varepsilon^{\,\nu}_{\mp}(k)=0
\end{displaymath}
as well as the closure relations
\begin{equation}
\varepsilon^{\,\mu\ast}_{+}(k)\,\varepsilon^{\,\nu}_{\,+}(k) +
\varepsilon^{\,\mu\ast}_{\,-}(k)\,\varepsilon^{\,\nu}_{-}(k) =
\varepsilon^{\,\mu}_{-}(k)\,\varepsilon^{\,\nu}_{\,+}(k) +
\varepsilon^{\,\mu}_{\,+}(k)\,\varepsilon^{\,\nu}_{-}(k)= {\mbox{\tt D}}^{-1}\,{S^{\,\mu\nu}}
\end{equation}

In order to obtain the normal modes expansion of the MCS
quantum field,
let's introduce the kinetic $4\times4$ hermitian kinetic matrix $\mathbb K$ with elements
\begin{equation}
K_{\,\lambda\nu}\equiv
g_{\,\lambda\nu}\left(k^2-m^2\right) +
i\varepsilon_{\lambda\nu\alpha\beta}\,\zeta^\alpha k^{\beta}
\end{equation}
which satisfies
\[
K_{\,\lambda\nu}=K^{\,\ast}_{\,\nu\lambda}
\]
Now we are ready to find the general solution of the free field equations
(\ref{MEL2}) .
As a matter of fact, from the relationships
(\ref{projectors1}) and (\ref{projectors2}) we readily obtain
\begin{eqnarray}
K^{\,\mu}_{\phantom{\mu}\nu}\,\varepsilon^{\,\nu}_{\pm}(k) &=&
\left[\,\delta^{\,\mu}_{\phantom{\mu}\nu}\left(k^2-m^{2}\right)
+ {\sqrt{\mbox{\tt D}}}\,\left(\,\bfpi^{\,\mu}_{\,+\,\nu}\;-\;\bfpi^{\,\mu}_{\,-\,\nu}\,\right)\,\right]
\varepsilon^{\,\mu}_{\pm}(k)\nonumber\\
&=& \left(k^2-m^{2} \pm\,\sqrt{\mbox{\tt D}} \,\right)\,
\varepsilon^{\,\mu}_{\pm}(k)
\end{eqnarray}
which shows that the polarization vectors of positive and negative chiralities
respectively are solutions of the vector field equations for $\zeta\cdot x<0$ iff
\begin{eqnarray}
k^{\,\mu}_{\,\pm}=(\omega_{\,{\bf k}\,,\,\pm}\,,\,{\bf k})\qquad\quad
\varepsilon^{\,\mu}_{\pm}({\bf k},\zeta_0)=\varepsilon^{\,\mu}_{\pm}(k_{\pm})\qquad\quad
\left(\,k^{0}_{\pm}\;=\;\omega_{\,{\bf k}\,,\,\pm}\,\right)\\
\omega_{\,{\bf k}\,,\,\pm}=\left\lbrace
\begin{array}{cc}
\sqrt{{\bf k}^2+m^{2}\pm\zeta_0 |\,{\bf k}\,|}
& {\rm for}\quad\zeta_\mu=(\zeta_0,0,0,0)\\
\sqrt{{\bf k}^2+m^{2}+\frac12\zeta_x^2\pm\zeta_x\sqrt{k_1^2+m^2+\frac14\zeta_x^2}}
& {\rm for}\quad\zeta_\mu=(0,-\,\zeta_x,0,0)
\end{array}\right.\label{MCSDR}
\end{eqnarray}
\\
It is worthwhile to enlighten on the above relationships with some comments and remarks.
Firstly, the four momentum $k^{\,\mu}_{-}=(\omega_{\,{\bf k}\,,\,-}\,,\,{\bf k})$,
that specifies the polarization vector $\varepsilon^{\,\mu}_{-}(k_{-})$
of negative chirality, turns out to stay
within the causal cone $k^2_->0$ iff
the spatial momentum is below an ultraviolet cutoff, i.e.,
$|\,{\bf k}\,|<m^2/|\zeta_0|$ in the temporal case or $|\,{\bf k}\,|<m^2/|\zeta_x|$ for the spatial
Chern-Simons vector \cite{AGS2002,AACGS2010,KL2001}.
Furthermore, the above birefringent dispersion relations fulfilled by
the chiral polarizations do admit quite different massless limits as discussed in \cite{AGS2002}.
As a matter of fact, in the spatial case $\zeta_\mu=(0,-\,\zeta_x,0,0)$ the massless limit
is smooth and safe since both angular frequencies $\omega_{\,{\bf k},\pm}$
keep real and the corresponding group velocities
$\nabla_{\mathbf k}\,\omega_{\,{\bf k},\pm}$
have modulus less than one in
natural units. Conversely, for the temporal case $\zeta_\mu=(\zeta_0,0,0,0)$ the massless
limit is troublesome because the angular frequency $\omega_{\,{\bf k},-}$
is imaginary in the infrared and both the related group velocities have modulus
which is always larger than one. Last but not least, it is crucial to realize that the chiral
polarizations of the Maxwell-Chern-Simons quanta have nothing to share with
the elliptic polarizations of the electromagnetic radiation fields since, for instance, the former undergo
vacuum birefringence, while the latter never do it.

To complete our construction of a basis
we suitably introduce the further pair of orthonormal polarization vectors,
respectively the so called scalar and longitudinal polarization real vectors
\begin{eqnarray}
\varepsilon^{\,\mu}_{\,S}(k)\;\equiv\;
\frac{k^{\,\mu}}{\sqrt{\,k^2}}\qquad\qquad
(\,k^2>0\,)\\
\varepsilon^{\,\mu}_{\,L}(k)\;\equiv\;\left(\mbox{\tt D}\,k^2\right)^{-\frac12}\left(
k^2\,\zeta^{\,\mu} - k^{\,\mu}\,\zeta\cdot k\,\right)
\qquad\qquad
(\,k^2>0\vee\mbox{\tt D}>0\,)
\end{eqnarray}
which fulfill by construction
\begin{eqnarray}
k_\mu\,\varepsilon^{\,\mu}_{\,L}(k)=0\qquad\quad
k_\mu\,\varepsilon^{\,\mu}_{\,S}(k)=\sqrt{k^2}\qquad(\,k^2>0\,)\\
g_{\mu\nu}\;\varepsilon^{\,\mu}_{\,S}(k)\,\varepsilon^{\,\nu}_{\,S}(k)\ =\ 1
\qquad\quad
g_{\mu\nu}\;\varepsilon^{\,\mu}_{\,L}(k)\,\varepsilon^{\,\nu}_{\,L}(k)\ =\ -\,1\\
g_{\,\mu\nu}\,\varepsilon^{\,\mu}_{\,S}(k)\,\varepsilon^{\,\nu}_{\,L}(k)=
g_{\,\mu\nu}\,\varepsilon^{\,\mu}_{\,S}(k)\,\varepsilon^{\,\nu}_{\pm}(k)=
g_{\,\mu\nu}\,\varepsilon^{\,\mu}_{\,L}(k)\,\varepsilon^{\,\nu}_{\pm}(k)=0
\end{eqnarray}
Hence we have at our disposal $\forall\,k^{\,\mu}$ with $k^2>0 \vee \mbox{\tt D}>0$ a complete
and orthonormal chiral set of four polarization vectors: namely,
\begin{eqnarray}
\varepsilon^{\,\mu}_{\,A}(k)=\left\lbrace
\begin{array}{cc}
{k^{\,\mu}}/{\sqrt{\,k^2}} & {\rm for}\ A=S\\
\left(k^2\,\zeta^{\,\mu} - k^{\,\mu}\,\zeta\cdot k\,\right)/
\displaystyle\sqrt{\mbox{\tt D}\,k^2} & {\rm for}\ A=L\\
\varepsilon^{\,\mu}_{\pm}(k_{\pm}) & {\rm for}\ A=\pm
\end{array}
\right.
\qquad\quad
(\,k^2>0\,\vee\,\mbox{\tt D}>0\,)
\end{eqnarray}
in such a manner that if we introduce the $4\times4$ polarization matrix
\begin{eqnarray}
g_{\,AB}=g^{\,AB}
\equiv\left\lgroup
\begin{array}{cccc}
1 & 0 & 0 & 0\\
0 & -1 & 0 & 0\\
0 & 0 & -1 & 0\\
0 & 0 & 0 & -1\
\end{array}\right\rgroup
\qquad\quad
(\, A,B=S,L,+,-\,)
\end{eqnarray}
then we can write the full orthogonality and closure relations
\begin{eqnarray}
g_{\mu\nu}\;\varepsilon^{\,\mu\ast}_{\,A}(k)\,\varepsilon^{\,\nu}_{\,B}(k)
\,=\,g_{AB}\qquad\quad
g^{AB}\,\varepsilon^{\,\mu\ast}_{\,A}(k)\,\varepsilon^{\,\nu}_{\,B}(k)
\,=\,g^{\,\mu\nu}
\label{orthoclosure}
\end{eqnarray}
together with the transversality relation
\begin{equation}
 k_{\nu}\,\varepsilon^{\,\nu}_{\,A}(k)=\sqrt{k^{2}}\,\delta_{AS}
\label{transversality}
\end{equation}
It is very important to keep in mind \cite{AACGS2010,AEGS2009} that the on mass shell polarization vector
$\varepsilon^{\,\mu}_{\,-}(k_-)$ with $k_-^2>0$
is well defined
iff the spatial momentum ${\bf k}$ keeps below the
momentum cutoff, i.e., inside the large momentum sphere
$|{\bf k}|<{m^2}/{\zeta}\equiv\Lambda$,
where $\zeta$ stands for either $|\zeta_0|$ or $|\zeta_x|$.

Now, in order to fully implement the canonical quantization of the MCS
massive vector field for the especially simple choice $\varkappa=1$, it is convenient to introduce
the polarized plane waves according to
\begin{eqnarray}
v_{\,{\bf k}\,A}^{\,\nu}(x) =
\left[\,(2\pi)^3\,2\omega_{\,{\bf k}\,A}\,\right]^{-\frac12}\,\varepsilon^{\,\nu}_{\,A}(k)\
\exp\{-\,i\omega_{\,{\bf k}\,A}\,x^0 + i{\bf k}\cdot{\bf x}\}
\end{eqnarray}
where the dispersion relation for the scalar and longitudinal frequencies is
the covariant one, viz.,
\[\omega_{\,{\bf k}\,S}=\omega_{\,{\bf k}\,L}=
\displaystyle\sqrt{{\bf k}^{2}+m^2}\equiv\omega_{\,{\bf k}}\]
so that we can write
\begin{equation}
 k_\nu\,\varepsilon^{\,\nu}_{\,S}(k)=m\qquad\quad
i\partial_\nu v_{\,{\bf k}\,S}^{\,\nu}(x)=u_{\,{\bf k}}(x)
\end{equation}
It follows therefrom that the general solution of the Euler-Lagrange equations (\ref{EL2}) for the quantized massive vector
field when $\varkappa=1$ and $\zeta\cdot x<0$ takes the form
\begin{eqnarray}
A^{\nu}(x) &=& A^{\nu}_{\rm CS}(x)-\partial^{\,\nu}B(x)/m^2\\
A^{\nu}_{\rm CS}(x)&=&
\int\mathrm d^3 \mathbf k\sum_{A=\pm,L}\,
\left[\,c_{\,{\bf k},A}\,v_{\,{\bf k}\,A}^{\,\nu}(x) +
c_{\,{\bf k},A}^{\,\dagger}\,v_{\,{\bf k}\,A}^{\,\nu\ast}(x)\,\right]\\
B(x)&=& {m}\int\mathrm d^3 \mathbf k\,
\left[\,b_{\,{\bf k}}\,u_{\,{\bf k}}(x) +
b_{\,{\bf k}}^{\,\dagger}\,u_{\,{\bf k}}^{\ast}(x)\,\right]
\end{eqnarray}
where the canonical commutation relations holds true, viz.,
\begin{eqnarray}
\left[\,c_{\,{\bf k},A}\,,\,c_{\,{\bf k}^{\prime},A^{\prime}}^{\,\dagger}\,\right]\;=\;-\,g_{AA^{\prime}}\,
\delta({\bf k}-{\bf k}^{\prime})\qquad\quad
c_{\,{\bf k},S}=b_{\,{\bf k}}
\end{eqnarray}
all the other commutators being equal to zero. According to equations (\ref{EL2}) and (\ref{MEL2})
we obtain
\begin{eqnarray}
B(x)= -\,i\int\mathrm d^3 \mathbf k\;k_{\nu}\left[\,c_{\,{\bf k}\,S}\,\,v_{\,{\bf k}\,S}^{\,\nu}(x) -
c_{\,{\bf k}\,S}^{\,\dagger}\,v_{\,{\bf k}\,S}^{\,\nu\ast}(x)\,\right]_{k_{0}\,=\,\omega_{\,{\bf k}}}=
\nonumber \\
=m\int\mathrm d^3 \mathbf k\;\left[\,b_{\,{\bf k}}\,u_{\,{\bf k}}(x) +
b_{\,{\bf k}}^{\,\dagger}\,u_{\,{\bf k}}^{\ast}(x)\,\right]_{k_{0}\,=\,\omega_{\,{\bf k}}}
\end{eqnarray}
in such a manner that the physical Hilbert space ${\mathfrak H}_{\,\rm phys}$ with positive semi-definite metric,
for the MCS massive quanta,
is selected out from the Fock space ${\mathfrak F}$ by means of the subsidiary condition
\begin{eqnarray}
B^{\,(-)}(x)\,|\,\rm phys\,\rangle\;=\;0\qquad\quad
\forall\,|\,\rm phys\,\rangle\,\in\,{\mathfrak H}_{\,\rm phys}\subset{\mathfrak F}
\end{eqnarray}
On the other side, it turns out that
the physical MCS massive quanta are created out of the Fock vacuum by the creation part of the quantized
physical massive MCS vector field $A^{\nu}_{\rm CS}(x)$,
with the standard nonvanishing canonical commutation relations
\begin{displaymath}
 \left[\,c_{\,{\bf k},A}\,,\,c_{\,{\bf k}^{\prime},A^{\prime}}^{\,\dagger}\,\right]\;=\;\delta_{AA^{\prime}}\,
\delta({\bf k}-{\bf k}^{\prime})\qquad\qquad A,A^{\prime}=L,\pm
\end{displaymath}
all the other commutators being equal to zero.
Notice that the MCS massive 1-particle states
of definite spatial momentum $\bf k$ do exhibit three polarization states, i.e.,
one linear longitudinal polarization of real vector $\varepsilon^{\,\nu}_{\,L}(k)$
with dispersion relation $k^2=m^2$
and two chiral transverse states with complex vectors $\varepsilon^{\,\nu}_{\,\pm}(k_{\pm})$
and dispersion relations (\ref{MCSDR}),
the negative chirality states $\varepsilon^{\,\nu}_{\,-}(k_-)$ being well defined
only for $|{\bf k}|<\Lambda\,\Leftrightarrow\,k_{-}^{2}>0$.
\section{The Bogolyubov Transformations}
The Proca-St\"{u}ckelberg vector field and the Maxwell-Chern-Simons massive vector field
face one another at the hyperplane $\zeta\cdot x=0$. Hence locality of the quantized wave fields does require
equality on the surface separating the classical pseudoscalar background from the vacuum: namely,
\begin{equation}
\delta(\zeta\cdot x)\left[ \,A^{\mu}_{\rm PS}(x)-A^{\mu}_{\rm CS}(x)\,\right] = 0 ,
\label{boundary}
\end{equation}
while the auxiliary unphysical field $B(x)$ is not at all affected by the presence of the
boundary $\zeta\cdot x=0$.
Let us first discuss the case of a spatial Chern-Simons vector $\zeta_{\mu}=(0,-\,\zeta_x,0,0)$ so that
$\delta(\zeta\cdot x)=\zeta_x^{-1}\delta(x)$
 and set such objects: $\hat k = (\omega, k_2, k_3)$, $\hat x = (x_0, x_2, x_3) $ : $\hat k \cdot \hat x =- \omega x_0 + k_2 x_2 + k_3 x_3$. We can write solution in form:
\begin{eqnarray}
A^{\mu}_{\rm PS}(x) &=& \int\mathrm d^3 \hat k \,\theta(\omega^2 - k^2_\bot - m^2)\sum_{r=1}^3\
\left[\,\mathbf{a}_{\,{\hat k}\,,\,r}\,u^{\,\mu}_{\,{\hat k}\,,\,r}(x)
+ \mathbf{a}^\dagger_{\,{\hat k}\,,\,r}\,
u^{\,\mu\,\ast}_{\,{\hat k}\,,\,r}(x)\,\right],\\
\partial_{\mu}A^{\mu}_{\rm PS}(x)&=&0 \nonumber \\
u^\nu_{{\hat k}\,,\,r}(x)&=&[\,(2\pi)^3\,2k_{10}\,]^{-1/2}\,
e_{\,r}^{\,\nu}({\hat k})\,
\exp\{\,i\,k_{10} x_1 + i\,{\hat k}\cdot{\hat x}\}\qquad (\,r=1,2,3\,)\nonumber\\
k_{10} &=& \sqrt{\omega^2 - k^2_\bot - m^2}
\end{eqnarray}
\\for $x_1>0$.
where the creation-destruction operators fulfill the canonical commutation relations
\begin{equation}
 [\,\mathbf a_{\,{\hat k}\,,\,r}\,,\,\mathbf a^\dagger_{\,{\hat k}^\prime\,,\,s}\,]=\delta({\hat k}-{\hat k}^\prime)\,\delta_{rs}
\qquad\quad
\end{equation}
all the remaining commutators being equal to zero.
The three linear polarization real vectors do satisfy the orthogonality and closure relations
on the mass shell $k^2=m^2\,:$ namely,
\begin{eqnarray}
k_{\mu}e_{\,r}^{\,\mu}({\hat k})=0\qquad
 -\,g_{\,\mu\nu}\;e_{\,r}^{\,\mu}({\hat k})\,
e_{\,s}^{\,\nu}({\hat k})=\delta_{\,rs}
\qquad
\sum_{r=1}^3\,e_{\,r}^{\,\mu}({\hat k})\,
e_{\,r}^{\,\nu}({\hat k})
= -\;g^{\;\mu\nu} + \frac{k^{\mu}k^{\nu}}{m^2}
\end{eqnarray}
And for $x_1<0$ we have:
\begin{eqnarray}
A^{\nu}_{\rm CS}(x)&=&
\int\mathrm d^3 \hat k \,\theta(\omega^2 - k^2_\bot - m^2)\sum_{A}\,
\left[\,c_{\,{\hat k},A}\,v_{\,{\hat k}\,A}^{\,\nu}(x) +
c_{\,{\hat k},A}^{\,\dagger}\,v_{\,{\hat k}\,A}^{\,\nu\ast}(x)\,\right]\\
A\in\{L,+,-\} \nonumber
\end{eqnarray}

\begin{eqnarray}
v_{\,{\hat k}\,A}^{\,\nu}(x) =
\left[\,(2\pi)^3\,2k_{1A}\,\right]^{-\frac12}\,\varepsilon^{\,\nu}_{\,A}(k)\
\exp\{\,i k_{10}\,x_1 + i{\hat k}\cdot{\hat x}\}
\end{eqnarray}

where the canonical commutation relations holds true, viz.,
\begin{eqnarray}
\left[\,c_{\,{\hat k},A}\,,\,c_{\,{\hat k}^{\prime},A^{\prime}}^{\,\dagger}\,\right]\;=\;-\,g_{AA^{\prime}}\,
\delta({\hat k}-{\hat k}^{\prime})\qquad\quad
\end{eqnarray}
all the other commutators being equal to zero.\\\\
So, boundary conditions become
\begin{eqnarray}
\nonumber
&&\int\mathrm d^3 \hat k \,\theta(\omega^2 - k^2_\bot - m^2)\times
\\
&&\!\!\!\!\!\!\times \left\lbrace \sum_{A}
\left[\,c_{\,{\hat k},A}\,v_{\,{\hat k},A}^{\,\mu}(\hat x) +
c_{\,{\hat k},A}^{\,\dagger}\,v_{\,{\hat k},A}^{\,\mu\ast}(\hat x)\,\right]
- \sum_{r=1}^3\
\left[\,\mathbf{a}_{\,{\hat k}\,,\,r}\,u^{\,\mu}_{\,{\hat k},r}(\hat x)
+ \mathbf{a}^\dagger_{\,{\hat k}\,,\,r}\,
u^{\,\mu\,\ast}_{\,{\hat k},r}(\hat x)\,\right]\right\rbrace =0
\label{boundary3}
\end{eqnarray}
Now, we suggest
 \begin{equation}
 v_{\,{\hat k},A}^{\,\nu}(\hat x)=\sum_{s=1}^3
 \,\left[\alpha_{s A}({\hat k})\,u^{\,\nu}_{\,{\hat k},s}(\hat x)-\beta_{s A}({\hat k})\,u^{\,\nu \ast}_{\,{\hat k},s}(\hat x) \right]
 \label{bg}
\end{equation}
On the one hand we can find
\begin{eqnarray}
\left\langle u^{\,\mu}_{\,{\hat p},r}\,|\,v_{\,{\hat k},A}^{\,\nu}\right\rangle
\equiv -i \int\mathrm d^3 \hat x\,
u^{\,\mu\ast}_{\,{\hat p},r}(t,y,z)\,\overleftrightarrow{\partial_1}v_{\,{\hat k},A}^{\,\nu}(t,y,z)\nonumber\\
=-\frac{k_{1A}+k_{10}}{2\sqrt{k_{1A} k_{10}}}\,
\exp\{ix_1(k_{10}-k_{1A})\}\,
\delta(\hat k-\hat p)\,e^{\mu}_{r}({\hat k})\,\varepsilon^{\nu}_{A}({\hat k})
\end{eqnarray}
On the other hand we obtain
\begin{eqnarray}
 \left\langle u^{\,\mu}_{\,{\hat p},r}\,|\,u_{\,{\hat k},s}^{\,\lambda}\right\rangle
=\delta(\hat k - \hat p)\,e^{\mu}_{r}({\hat k})\,e^{\lambda}_{s}({\hat k})
\end{eqnarray}
\begin{equation}
\left\langle u^{\,\mu}_{\,{\hat p},r}\,|\,u_{\,{\hat k},s}^{\,\lambda\ast}\right\rangle=0
\end{equation}
and thereby
\begin{eqnarray}
 \left\langle u^{\,\mu}_{\,{\hat p},r}\,|\,v_{\,{\hat k},A}^{\,\nu}\right\rangle
= \delta(\hat k-\hat p)\sum_{s=1}^3 \alpha_{s A}({\hat k})\,e^{\nu}_{s}({\hat k})\,e^{\mu}_{r}({\hat k})\\
\left\langle u^{\,\mu\ast}_{\,{\hat p},r}\,|\,v_{\,{\hat k},A}^{\,\nu}\right\rangle
= 0
\end{eqnarray}
A comparison yields
\begin{equation}
 -\frac{k_{1A}+k_{10}}{2\sqrt{k_{1A}k_{10}}}\,
\exp\{ix_1(k_{10}-k_{1A})\}\,\varepsilon^{\nu}_{A}({\hat k})=
\sum_{s=1}^3 \alpha_{s A}(t,{\hat k})\,e^{\nu}_{s}({\hat k})
\end{equation}
the solution of which is provided by
\begin{equation}
 \alpha_{s A}({\hat k})=
-\,{\textstyle\frac12}\,g_{\mu\nu}\,\varepsilon^{\,\mu}_{A}({\hat k})\,e^{\nu}_{s}({\hat k})\;
\frac{k_{1A}+k_{10}}{\sqrt{k_{1A}k_{10}}}\,
\exp\{ix_1(k_{10}-k_{1A})\}
\end{equation}
But for our boundary $x_1=0$, so $\exp\{ix_1(k_{10}-k_{1A})\}=1$ and we can write
\begin{equation}
 \alpha_{s A}({\hat k})=
-\,{\textstyle\frac12}\,g_{\mu\nu}\,\varepsilon^{\,\mu}_{A}({\hat k})\,e^{\nu}_{s}({\hat k})\;
\frac{k_{1A}+k_{10}}{\sqrt{k_{1A}k_{10}}}
\end{equation}
as it can be readily checked by direct substitution. \\
Let's evaluate the matrix elements
\begin{displaymath}
\sum_{s=1}^3\left[\, \alpha_{s A}({\hat k})\,\alpha_{s B}^{\ast}({\hat k})
-\beta_{s A}({\hat k})\,\beta^{\ast}_{s B}({\hat k})\,\right]
\qquad\qquad A,B=L,\pm
\end{displaymath}
First we find
\begin{eqnarray*}
&&{\textstyle\frac12}\,g_{\mu\nu}\,\varepsilon^{\,\mu}_{A}({\hat k})\,
{\textstyle\frac12}\,g_{\iota\kappa}\,\varepsilon^{\,\iota\ast}_{B}({\hat k})
\sum_{s=1}^3 e^{\nu}_{s}({\hat k})\,\,e^{\kappa}_{s}({\hat k})\\
&=&{\textstyle\frac14}\,
\varepsilon^{\,\mu}_{A}({\hat k})\,\varepsilon^{\,\iota\ast}_{B}({\hat k})
\left( -\,g_{\iota\mu} + \frac{k_{\iota}k_{\mu}}{m^{2}}\right)
= -\,{\textstyle\frac14}\,g_{\iota\mu}\,
\varepsilon^{\,\mu}_{A}({\hat k})\,\varepsilon^{\,\iota\ast}_{B}({\hat k})
= {\textstyle\frac14}\,\delta_{AB}
\end{eqnarray*}
where use has been made of the transversality condition (\ref{transversality}), as well as
the orthonormality relations (\ref{orthonorma}), together with
the fact that the covariant linear polarization vectors have been chosen to be real.
Hence we eventually obtain
\begin{eqnarray*}
 \sum_{s=1}^3 \alpha_{s A}({\hat k})\,\alpha_{s B}^{\ast}({\hat k})=
\delta_{AB}\;\frac{(k_{1A}+k_{10})^2}
{4k_{1A}k_{10}}\\
\sum_{s=1}^3 \beta_{s A}({\hat k})\,\beta^{\ast}_{s B}({\hat k})=
\delta_{AB}\;\frac{(k_{1A}-k_{10})^{2}}
{4k_{1A}k_{10}}
\end{eqnarray*}
Subtraction of the above expressions yields the customary relation
\begin{equation}
\sum_{s=1}^3\left[\, \alpha_{s A}({\hat k})\,\alpha_{s B}^{\ast}({\hat k})
-\beta_{s A}({\hat k})\,\beta^{\ast}_{s B}({\hat k})\,\right]=\delta_{AB}
\qquad\qquad A,B=L,\pm
\end{equation}
and by making quite analogous manipulations one can readily check that the further
usual Bogolyubov relations
\begin{equation}
\sum_{s=1}^3\left[\, \alpha_{s A}({\hat k})\,\beta_{s B}^{\ast}({\hat k})
-\beta_{s A}({\hat k})\,\alpha^{\ast}_{s B}({\hat k})\,\right]=0
\qquad\qquad A,B=L,\pm
\end{equation}
\begin{equation}
\sum_{s=1}^3\left[\, \alpha_{s A}({\hat k})\,\beta_{s B}({\hat k})
-\beta_{s B}({\hat k})\,\alpha_{s A}({\hat k})\,\right]=0
\qquad\qquad A,B=L,\pm
\end{equation}

Turning back to the boundary condition (\ref{boundary3}) and taking the Bogolyubov transformation
(\ref{bg}) into account, we can write the operator equalities
\begin{eqnarray}
\label{relation}
\mathbf a_{\,{\hat k},\,r}=\sum_{A=\pm,L}\left[ \,\alpha_{\,r A}({\hat k})\,c_{\,{\hat k},A}
- \beta^{\ast}_{\,r A}({\hat k})\,c_{\,{\hat k},A}^{\,\dagger}\,\right]\\
c_{\,{\hat k},A}=\sum_{r=1}^3\left[ \,\alpha^{\ast}_{A r}({\hat k})\,\mathbf a_{\,{\hat k},\,r}
+ \beta^{\ast}_{A r}({\hat k})\,\mathbf a_{\,{\hat k},\,r}^{\,\dagger}\,\right]
\label{yubov1}
\end{eqnarray}
From the canonical commutation relations we obtain
\begin{eqnarray*}
 \left[ \,\mathbf a_{\,{\hat k},\,r}\,,\,\mathbf a^{\dagger}_{\,{\hat p},\,s}\,\right]
= \delta({\hat k}-{\hat p})\,\delta_{rs}\\
=\sum_{A,B=\pm,L}\left[ \,\alpha_{\,r A}({\hat k})\,c_{\,{\hat k},A}
- \beta^{\ast}_{\,r A}({\hat k})\,c_{\,{\hat k},A}^{\,\dagger}\,,\,
\alpha^{\ast}_{\,s B}({\hat p})\,c^{\,\dagger}_{\,{\hat p},B}
- \beta_{\,s B}({\hat p})\,c_{\,{\hat p},B}\,\right]\\
=\sum_{A=\pm,L}\left[\, \alpha_{\,r A}({\hat k})\,\alpha^{\ast}_{\,s A}({\hat p})
- \beta_{\,r A}({\hat k})\,\beta^{\ast}_{\,s A}({\hat p})\,\right]
\delta({\hat k}-{\hat p})
\end{eqnarray*}
and consequently
\begin{equation}
 \sum_{A=\pm,L}\left[\, \alpha_{\,r A}({\hat k})\,\alpha^{\ast}_{\,s A}({\hat k})
- \beta_{\,r A}({\hat k})\,\beta^{\ast}_{\,s A}({\hat k})\,\right]=\delta_{rs}
\end{equation}
The null commutators
$\left[ \,a_{\,{\hat k},\,r}\,,\,a_{\,{\hat p},\,s}\,\right]=
\left[ \,a^{\dagger}_{\,{\hat k},\,r}\,,\,a^{\dagger}_{\,{\hat p},\,s}\,\right]=0$
lead to the further relations
\begin{equation}
 \sum_{A=\pm,L}\left[\, \alpha_{\,r A}({\hat k})\,\beta^{\ast}_{\,s A}({\hat k})
- \beta_{\,r A}({\hat k})\,\alpha^{\ast}_{\,s A}({\hat k})\,\right]=0
\end{equation}
\begin{equation}
 \sum_{A=\pm,L}\left[\, \alpha_{\,r A}({\hat k})\,\beta_{\,s A}({\hat k})
- \beta_{\,r A}({\hat k})\,\alpha_{\,s A}({\hat k})\,\right]=0
\end{equation}
There are two different Fock vacua: namely,
\begin{displaymath}
\mathbf a_{\,{\hat k},\,r}\vert0\rangle=0 \qquad\quad
c_{\,{\hat k},A}\mid\Omega\,\rangle=0
\end{displaymath}
whence
\[c_{\,{\hat k},A}\mid0\,\rangle=\sum_{r=1}^3\beta^{\ast}_{A r}({\hat k})\,a_{\,{\hat k},\,r}^{\,\dagger}\mid0\,\rangle\]
and consequently
\begin{eqnarray*}
 \langle\,0\mid c_{\,{\hat p},B}^{\,\dagger}\,c_{\,{\hat k},A}\mid0\,\rangle&=&
\sum_{r,s=1}^3\langle\,0\mid \mathbf a_{\,{\hat p},s}\,\mathbf a_{\,{\hat k},\,r}^{\,\dagger}\mid0\,\rangle\,
\beta_{B s}({\hat p})\,\beta^{\ast}_{A r}({\hat k})\\
&=&\delta({\hat k}-{\hat p})\sum_{r=1}^3\beta_{B r}({\hat k})\,\beta^{\ast}_{A r}({\hat k})
=\delta({\hat k}-{\hat p})\,\delta_{AB}\;\frac{(k_{1A}-k_{10})^{2}}
{4k_{1A}k_{10}}
\end{eqnarray*}
In turn we evidently obtain
\begin{equation}
 \mathbf a_{\,{\hat k},\,r}\mid\Omega\,\rangle =
-\sum_{A=L,\pm}\beta^{\ast}_{\,r A}({\hat k})\,c_{{\hat k},A}^{\,\dagger}\mid\Omega\,\rangle
\end{equation}
that yields
\begin{eqnarray*}
 \langle\,\Omega\mid \mathbf a_{\,{\hat p},s}^{\,\dagger}\,\mathbf a_{\,{\hat k},r}\mid\Omega\,\rangle&=&
\sum_{A,B=L,\pm}\langle\,\Omega\mid c_{\,{\hat p},A}\,c_{\,{\hat k},\,B}^{\,\dagger}\mid\Omega\,\rangle\,
\beta_{B s}({\hat p})\,\beta^{\ast}_{A r}({\hat k})\\
&=&\delta({\hat k}-{\hat p})\sum_{A=L,\pm}\beta_{A s}({\hat k})\,\beta^{\ast}_{A r}({\hat k})
\end{eqnarray*}
Moreover we get
\begin{eqnarray*}
 \langle\,0\mid \mathbf a_{\,{\hat p},s}\,c_{\,{\hat k},A}\mid0\,\rangle
=\langle\,0\mid \mathbf a_{\,{\hat p},s}\,\sum_{r=1}^3\left[ \,\alpha^{\ast}_{A r}({\hat k})\,\mathbf a_{\,{\hat k},\,r}
+ \beta^{\ast}_{A r}({\hat k})\,\mathbf a_{\,{\hat k},\,r}^{\,\dagger}\,\right]\mid0\,\rangle\\
=\sum_{r=1}^3\beta^{\ast}_{A r}({\hat k})\,\langle\,0\mid \mathbf a_{\,{\hat p},s}\,\mathbf a^{\,\dagger}_{\,{\hat k},\,r}\mid0\,\rangle
=\delta({\hat k}-{\hat p})\,\beta^{\ast}_{A s}({\hat k})\\
\langle\,0\mid \mathbf a_{\,{\hat p},s}\,c^{\,\dagger}_{\,{\hat k},A}\mid0\,\rangle
=\langle\,0\mid \mathbf a_{\,{\hat p},s}\,\sum_{r=1}^3\left[ \,\alpha_{A r}({\hat k})\,\mathbf a^{\,\dagger}_{\,{\hat k},\,r}
+ \beta_{A r}({\hat k})\,\mathbf a_{\,{\hat k},\,r}\,\right]\mid0\,\rangle\\
=\sum_{r=1}^3\alpha_{A r}({\hat k})\,\langle\,0\mid \mathbf a_{\,{\hat p},s}\,\mathbf a^{\,\dagger}_{\,{\hat k},\,r}\mid0\,\rangle
=\delta({\hat k}-{\hat p})\,\alpha_{A s}({\hat k})
\end{eqnarray*}
The latter quantity $\alpha_{A s}({\hat k})$ can thereof be interpreted as the
relative probability amplitude that a birefringent
particle of mass $m$, frequency  $\omega$ and wave vector $(k_{1A},k_2,k_3)$ and chiral polarization vector $\varepsilon^{\,\mu}_{A}({\hat k})$
is transmitted from the left face to the right face through the hyperplane $ x_1=0$ to become
a Proca-St\"uckelberg particle with equal mass $m$, frequency $\omega$ and wave vector $(k_{10}, k_2, k_3)$,
but polarization vector $e^{\,\mu}_{s}({\hat k})$. As an effect of this transmission, the first component of wave vector of a
birefringent massive particle
changes from $k_{1\pm}$ to $k_{10}$, while the longitudinal massive quanta do
not change it's wave vector.

\section{CS Electrodynamics for a space-like CS vector}
\subsection{Classical solutions}
Let us consider in more details the case of a spatial Chern-Simons vector $\zeta_{\mu}=(0,-\,\zeta_x,0,0)$  in the Lagrange density \eqref{lagrangian2}.
After eliminating the auxiliary field one obtains the classical equations \eqref{EL2}
 which in components read,
\begin{eqnarray}
\left\lbrace
\begin{array}{cc}
(\Box + m^2) A^0=-\zeta_x \theta(-x_1) (\partial_2 A_3 - \partial_3 A_2)\\
(\Box + m^2) A^1=0\\
(\Box + m^2) A^2=-\zeta_x  \theta(-x_1) (\partial_3 A_0 - \partial_0 A_3)\\
(\Box + m^2) A^3=-\zeta_x \theta(-x_1) (\partial_0 A_2 - \partial_2 A_0)
\end{array}\right.\label{MaxCS}
\end{eqnarray}
Let's introduce the vectors: $\hat k = (\omega, k_2, k_3)$, $\hat x = (x_0, x_2, x_3) $, and their scalar product $\hat k \cdot \hat x =- \omega x_0 + k_2 x_2 + k_3 x_3$. Using the Fourier transformation in these coordinates one can solve the equation describing $A_1$ in the entire space,
\begin{eqnarray}
A_1 = \int  \frac{d^3 \hat k}{(2\pi)^3}\ \theta(\omega^2 - k^2_\bot - m^2) \!\ (\tilde u_{1 \rightarrow}(\omega,k_2,k_3) e^{ik_{10} x_1}+\tilde u_{1 \leftarrow}(\omega,k_2,k_3) e^{-ik_{10} x_1}) e^{i \hat k \hat x}
\end{eqnarray}
where\\
$k_{10}^2 = \omega^2-m^2-k_\bot^2,\ k_\bot^2 = k_2^2 + k_3^2$.\\
Now, let's examine the remaining components of $A_\mu$. There are two solutions for each component in the different half-spaces. The first solution  describes the vector meson physics at $x_1>0$, the second one  is for $x_1<0$.\\
In the case $x_1>0$ one employs the solutions of the Maxwell equations without CS term on the entire axis,
\begin{eqnarray}
\left\lbrace
\begin{array}{cc}
A_0 = \int  \frac{\mathrm d^3 \hat k}{(2\pi)^3}\ \theta(\omega^2 - k^2_\bot - m^2) \!\ (\tilde u_{0 \rightarrow}(\omega,k_2,k_3) e^{ik_{10} x_1}+\tilde u_{0 \leftarrow}(\omega,k_2,k_3) e^{-ik_{10} x_1}) e^{i \hat k \hat x}\\
A_2 = \int  \frac{\mathrm d^3 \hat k}{(2\pi)^3}\ \theta(\omega^2 - k^2_\bot - m^2) \!\ (\tilde u_{2 \rightarrow}(\omega,k_2,k_3) e^{ik_{10} x_1}+\tilde u_{2 \leftarrow}(\omega,k_2,k_3) e^{-ik_{10} x_1}) e^{i \hat k \hat x}\\
A_3 = \int  \frac{\mathrm d^3 \hat k}{(2\pi)^3}\ \theta(\omega^2 - k^2_\bot - m^2) \!\ (\tilde u_{3 \rightarrow}(\omega,k_2,k_3) e^{ik_{10} x_1}+\tilde u_{3 \leftarrow}(\omega,k_2,k_3) e^{-ik_{10} x_1}) e^{i \hat k \hat x}\\
\end{array}\right.
\end{eqnarray}
To solve Eqs. \eqref{MaxCS} in the second case, we perform the Fourier transformation in $x_1$ as well on the entire axis,
\begin{eqnarray}
\left\lbrace
\begin{array}{cc}
(-\omega^2+\textbf{k}^2+m^2)\tilde{A_0}= - i\zeta_x  (k_3 \tilde{A_2} - k_2\tilde{A_3})\\
(-\omega^2+\textbf{k}^2+m^2)\tilde{A_2}=- i \zeta_x  (k_3 \tilde{A_0} + \omega\tilde{A_3})\\
(-\omega^2+\textbf{k}^2+m^2)\tilde{A_3}= i \zeta_x  (\omega \tilde{A_2} + k_2\tilde{A_0}) \label{system}
\end{array}\right.
\end{eqnarray}
This system leads to,
 \begin{eqnarray}
\tilde A_\nu=\sum_A\left[\tilde v_{\nu A\rightarrow}(k_2, k_3, \omega) \delta (k_1 - k_{1A})+\tilde v_{\nu A\leftarrow}(k_2, k_3, \omega) \delta (k_1 + k_{1A})\right]
\end{eqnarray}
Herein the first index of $\tilde v$ denotes  the corresponding component of $A_\nu$, $\nu=0,2,3$, the second index $A$ stays for different mass-shell $k_1$ for polarizations $L, +, -$ and the arrows $\rightarrow$, $\leftarrow$ indicate the direction of particle propagation. The dispersion laws for different polarizations read,
\begin{eqnarray}
\left\lbrace
\begin{array}{cc}
k_{1L}=k_{10}=\sqrt{\omega^2-m^2-k_\bot^2}\\
k_{1+}=\sqrt{\omega^2-m^2-k_\bot^2+\zeta_x \sqrt{\omega^2-k_\bot^2}}\\
k_{1-}=\sqrt{\omega^2-m^2-k_\bot^2-\zeta_x \sqrt{\omega^2-k_\bot^2}}\\
\end{array}\right.
\end{eqnarray}
As well there are following relations between $v$,
\begin{eqnarray}
\left\lbrace
\begin{array}{cc}
\tilde v_{2+\leftrightarrows}=\frac{k_2k_3-i\omega\sqrt{\omega^2-k_\bot^2}}{\omega^2-k_2^2}\tilde v_{3+\leftrightarrows}\\
\tilde v_{2-\leftrightarrows}=\frac{k_2k_3+i\omega\sqrt{\omega^2-k_\bot^2}}{\omega^2-k_2^2}\tilde v_{3-\leftrightarrows}\\
\\
\tilde v_{0+\leftrightarrows}=-\frac{i(k_3\tilde v_{2+\leftrightarrows}-k_2\tilde v_{3+\leftrightarrows})}{\sqrt{\omega^2-k_\bot^2}}\\
\tilde v_{0-\leftrightarrows}=\frac{i(k_3\tilde v_{2-\leftrightarrows}-k_2\tilde v_{3-\leftrightarrows})}{\sqrt{\omega^2-k_\bot^2}}\\
\\
\tilde v_{2L\leftrightarrows}=\frac{k_2}{k_3}\tilde v_{3L\leftrightarrows}\\
\tilde v_{0L\leftrightarrows}=-\frac{\omega}{k_3}\tilde v_{3L\leftrightarrows}\\
\end{array}\right.
\label{v}
\end{eqnarray}
Now we have the solutions in both parts of the space, and we have to match them on the boundary. For this purpose, let's take the system (\ref{system}) and perform Fourier transformation, using $x_0$, $x_2$, $x_3$.
\begin{eqnarray}
\left\lbrace
\begin{array}{cc}
(-\omega^2 + m^2 + k_\bot^2)\tilde A_0 - \partial_1^2 \tilde A_0= i \zeta_x \theta(-x_1)(k_2\tilde A_3 - k_3 \tilde A_2)\\
(-\omega^2 + m^2 + k_\bot^2)\tilde A_2 - \partial_1^2 \tilde A_2= -i \zeta_x \theta(-x_1)(k_3\tilde A_0 + \omega \tilde A_3)\\
(-\omega^2 + m^2 + k_\bot^2)\tilde A_3 - \partial_1^2 \tilde A_3= i \zeta_x \theta(-x_1)(\omega \tilde A_2 + k_2 \tilde A_0)\\
\end{array}\right.
\label{systFourier}
\end{eqnarray}
This system is valid in all the space. To solve it we have to integrate over $x_1$ in a small vicinity of the boundary, from $-\varepsilon$ to $\varepsilon$ and compare the coefficients of the exponentials in the left and right sides of the equations.
The solutions($\nu=0,2,3$) have the following general form,
\begin{eqnarray}
\tilde A_\nu = \left\lbrace
\begin{array}{cc}
\tilde u_{\nu \rightarrow}(\omega,k_2,k_3)e^{ik_{10} x_1}+\tilde u_{\nu \leftarrow}(\omega,k_2,k_3)e^{-ik_{10} x_1},\!\ x_1>0\\\\
\sum_A \left[\tilde v_{\nu A \rightarrow}(\omega,k_2,k_3) e^{ik_{1A} x_1}+\tilde v_{\nu A \leftarrow}(\omega,k_2,k_3) e^{-ik_{1A} x_1}\right], \!\ x_1<0
\end{array} \right. \label{gensol}
\end{eqnarray}
After careful integration of the boundary vicinity performed in Appendix A
we get additional relations from boundary conditions:
\begin{eqnarray}
\!\!\!\!\!\!\!\!\!\!\left\lbrace
\begin{array}{cc}
-k_{10}^2(-\frac{\tilde u_{0 \rightarrow}-\tilde u_{0 \leftarrow}}{i k_{10}}+\sum_{A}\frac{\tilde v_{0A\rightarrow}-\tilde v_{0A\leftarrow}}{i k_{1A}})= i \zeta_x (k_2\sum_{A}\frac{\tilde v_{3A\rightarrow}-\tilde v_{3A\leftarrow}}{i k_{1A}}-k_3\sum_{A}\frac{\tilde v_{2A\rightarrow}-\tilde v_{2A\leftarrow}}{i k_{1A}})\\
\\
-k_{10}^2(-\frac{\tilde u_{2 \rightarrow}-\tilde u_{2 \leftarrow}}{i k_{10}}+\sum_{A}\frac{\tilde v_{2A\rightarrow}-\tilde v_{2A\leftarrow}}{i k_{1A}})= - i \zeta_x (k_3\sum_{A}\frac{\tilde v_{0A\rightarrow}-\tilde v_{0A\leftarrow}}{i k_{1A}} + \omega\sum_{A}\frac{\tilde v_{3A\rightarrow}-\tilde v_{3A\leftarrow}}{i k_{1A}})\\
\\
-k_{10}^2(-\frac{\tilde u_{3 \rightarrow}-\tilde u_{3 \leftarrow}}{i k_{10}}+\sum_{A}\frac{\tilde v_{3A\rightarrow}-\tilde v_{3A\leftarrow}}{i k_{1A}})= i \zeta_x (\omega\sum_{A}\frac{\tilde v_{2A\rightarrow}-\tilde v_{2A\leftarrow}}{i k_{1A}} + k_2\sum_{A}\frac{\tilde v_{0A\rightarrow}-\tilde v_{0A\leftarrow}}{i k_{1A}})\\
\end{array}\right.
\end{eqnarray}
The contributions to amplitude in the right half-space ($u_\mu$) from the different dispersion laws in the left half-space are independent: $u_\mu=u_\mu^{(L)}+u_\mu^{(+)}+u_\mu^{(-)}$.\\
If $A=L$:
\begin{eqnarray}
\nonumber
\left\lbrace
\begin{array}{cc}
-k_{10}^2(-\frac{\tilde u_{0 \rightarrow}^{(L)}-\tilde u_{0 \leftarrow}^{(L)}}{i k_{10}}+\frac{\tilde v_{0L\rightarrow}-\tilde v_{0L\leftarrow}}{i k_{10}})= i \zeta_x (k_2\frac{\tilde v_{3L\rightarrow}-\tilde v_{3L\leftarrow}}{i k_{10}}-k_3\frac{\tilde v_{2L\rightarrow}-\tilde v_{2L\leftarrow}}{i k_{10}})\\
\\
-k_{10}^2(-\frac{\tilde u_{2 \rightarrow}^{(L)}-\tilde u_{2 \leftarrow}^{(L)}}{i k_{10}}+\frac{\tilde v_{2L\rightarrow}-\tilde v_{2L\leftarrow}}{i k_{10}})= - i \zeta_x (k_3\frac{\tilde v_{0L\rightarrow}-\tilde v_{0L\leftarrow}}{i k_{10}} + \omega\frac{\tilde v_{3L\rightarrow}-\tilde v_{3L\leftarrow}}{i k_{10}})\\
\\
-k_{10}^2(-\frac{\tilde u_{3 \rightarrow}^{(L)}-\tilde u_{3 \leftarrow}^{(L)}}{i k_{10}}+\frac{\tilde v_{3L\rightarrow}-\tilde v_{3L\leftarrow}}{i k_{10}})= i \zeta_x (\omega\frac{\tilde v_{2L\rightarrow}-\tilde v_{2L\leftarrow}}{i k_{10}} + k_2\frac{\tilde v_{0L\rightarrow}-\tilde v_{0L\leftarrow}}{i k_{10}})\\
\end{array}\right.
\end{eqnarray}
It is easy to see that the right parts in this system are equal to zero (from \eqref{v}), so we get
\begin{eqnarray}
\tilde u_{\nu \rightarrow}^{(L)}-\tilde u_{\nu \leftarrow}^{(L)}=\tilde v_{\nu L \rightarrow} - \tilde v_{\nu L \leftarrow}
\end{eqnarray}\\
If $A=+$:
\begin{eqnarray}
\nonumber
\left\lbrace
\begin{array}{cc}
-k_{10}^2(-\frac{\tilde u_{0 \rightarrow}^{(+)}-\tilde u_{0 \leftarrow}^{(+)}}{i k_{10}}+\frac{\tilde v_{0+\rightarrow}-\tilde v_{0+\leftarrow}}{i k_{1+}})= i \zeta_x (k_2\frac{\tilde v_{3+\rightarrow}-\tilde v_{3+\leftarrow}}{i k_{1+}}-k_3\frac{\tilde v_{2+\rightarrow}-\tilde v_{2+\leftarrow}}{i k_{1+}})\\
\\
-k_{10}^2(-\frac{\tilde u_{2 \rightarrow}^{(+)}-\tilde u_{2 \leftarrow}^{(+)}}{i k_{10}}+\frac{\tilde v_{2+\rightarrow}-\tilde v_{2+\leftarrow}}{i k_{1+}})= - i \zeta_x (k_3\frac{\tilde v_{0+\rightarrow}-\tilde v_{0+\leftarrow}}{i k_{1+}} + \omega\frac{\tilde v_{3+\rightarrow}-\tilde v_{3+\leftarrow}}{i k_{1+}})\\
\\
-k_{10}^2(-\frac{\tilde u_{3 \rightarrow}^{(+)}-\tilde u_{3 \leftarrow}^{(+)}}{i k_{10}}+\frac{\tilde v_{3+\rightarrow}-\tilde v_{3+\leftarrow}}{i k_{1+}})= i \zeta_x (\omega\frac{\tilde v_{2+\rightarrow}-\tilde v_{2+\leftarrow}}{i k_{1+}} + k_2\frac{\tilde v_{0+\rightarrow}-\tilde v_{0+\leftarrow}}{i k_{1+}})\\
\end{array}\right.
\end{eqnarray}
using the relations between $v_+$ \eqref{v}:
\begin{eqnarray}
\nonumber
\left\lbrace
\begin{array}{cc}
-k_{10}^2(-\frac{\tilde u_{0 \rightarrow}^{(+)}-\tilde u_{0 \leftarrow}^{(+)}}{i k_{10}}+\frac{\tilde v_{0+\rightarrow}-\tilde v_{0+\leftarrow}}{i k_{1+}})=(\tilde v_{0+\rightarrow}-\tilde v_{0+\leftarrow}) \frac{\zeta_x \sqrt{\omega^2+k_\bot^2}}{i k_{1+}} \\
\\
-k_{10}^2(-\frac{\tilde u_{2 \rightarrow}^{(+)}-\tilde u_{2 \leftarrow}^{(+)}}{i k_{10}}+\frac{\tilde v_{2+\rightarrow}-\tilde v_{2+\leftarrow}}{i k_{1+}})=(\tilde v_{2+\rightarrow}-\tilde v_{2+\leftarrow}) \frac{\zeta_x \sqrt{\omega^2+k_\bot^2}}{i k_{1+}}\\
\\
-k_{10}^2(-\frac{\tilde u_{3 \rightarrow}^{(+)}-\tilde u_{3 \leftarrow}^{(+)}}{i k_{10}}+\frac{\tilde v_{3+\rightarrow}-\tilde v_{3+\leftarrow}}{i k_{1+}})=(\tilde v_{3+\rightarrow}-\tilde v_{3+\leftarrow}) \frac{\zeta_x \sqrt{\omega^2+k_\bot^2}}{i k_{1+}}\\
\end{array}\right.
\end{eqnarray}
and for $A=+$ we get:
\begin{eqnarray}
\tilde u_{\nu \rightarrow}^{(+)}-\tilde u_{\nu \leftarrow}^{(+)}=\frac{(\tilde v_{\nu + \rightarrow} - \tilde v_{\nu + \leftarrow})k_{1+}}{k_{10}}
\end{eqnarray}
Finally, if $A=-$ using the same method we obtain:
\begin{eqnarray}
\tilde u_{\nu \rightarrow}^{(-)}-\tilde u_{\nu \leftarrow}^{(-)}=\frac{(\tilde v_{\nu - \rightarrow} - \tilde v_{\nu - \leftarrow})k_{1-}}{k_{10}}
\end{eqnarray}
Thus we have found the following matching conditions,
\begin{eqnarray}
\begin{array}{cc}
\tilde u_{\nu \rightarrow}^{(L)}-\tilde u_{\nu \leftarrow}^{(L)}=\tilde v_{\nu L \rightarrow} - \tilde v_{\nu L \leftarrow}\\
\tilde u_{\nu \rightarrow}^{(+)}-\tilde u_{\nu \leftarrow}^{(+)}=\frac{(\tilde v_{\nu + \rightarrow} - \tilde v_{\nu + \leftarrow})k_{1+}}{k_{10}}\\
\tilde u_{\nu \rightarrow}^{(-)}-\tilde u_{\nu \leftarrow}^{(-)}=\frac{(\tilde v_{\nu - \rightarrow} - \tilde v_{\nu - \leftarrow})k_{1-}}{k_{10}}
\end{array}
\end{eqnarray}
In addition, all contributions of $A$ are continuous.
\begin{eqnarray}
\begin{array}{cc}
\tilde u_{\nu \rightarrow}^{(L)}+\tilde u_{\nu \leftarrow}^{(L)}=\tilde v_{\nu L \rightarrow} + \tilde v_{\nu L \leftarrow}\\
\tilde u_{\nu \rightarrow}^{(+)}+\tilde u_{\nu \leftarrow}^{(+)}=\tilde v_{\nu + \rightarrow} + \tilde v_{\nu + \leftarrow}\\
\tilde u_{\nu \rightarrow}^{(-)}+\tilde u_{\nu \leftarrow}^{(-)}=\tilde v_{\nu - \rightarrow} + \tilde v_{\nu - \leftarrow}
\end{array}
\end{eqnarray}
From these sets of equations we obtain the relations between $u$ and $v$. All of them can be written in such a form,
\begin{eqnarray}
\tilde u_{\nu \rightarrow}^{(A)}=\frac{1}{2}(\tilde v_{\nu A \rightarrow}(\frac{k_{1A}+k_{10}}{k_{10}}) - \tilde v_{\nu A \leftarrow}(\frac{k_{1A}-k_{10}}{k_{10}}))\\
\tilde u_{\nu \leftarrow}^{(A)}=\frac{1}{2}(-\tilde v_{\nu A \rightarrow}(\frac{k_{1A}-k_{10}}{k_{10}}) + \tilde v_{\nu A \leftarrow}(\frac{k_{1A}+k_{10}}{k_{10}}))
\label{uv}
\end{eqnarray}\\
\subsection {Comparison of different representations}
Let's compare our classical solutions for a space-like CS vector with the solutions obtained by quantization formalism for the Bogolyubov transformation in Sec. 3. They must be the same, consequently we have,
\begin{eqnarray}
\frac{\tilde u_{\nu \rightarrow}}{(2\pi)^3} = \sum_{r=1}^{3} [\,(2\pi)^3\,2k_{10}\,]^{-1/2}\,
\mathbf a_{\hat k, r} e_{\,r}^{\,\nu}({\hat k})\,\\
\frac{\tilde v_{\nu A \rightarrow}}{(2\pi)^3} =  [\,(2\pi)^3\,2k_{1A}\,]^{-1/2}\,
\mathbf c_{\hat k, A} \varepsilon_{\,A}^{\,\nu}({\hat k})\,\\
\frac{\tilde v_{\nu A \leftarrow}}{(2\pi)^3} =  [\,(2\pi)^3\,2k_{1A}\,]^{-1/2}\,
\mathbf c_{\hat k, A}^\dag \varepsilon_{\,A}^{\,\nu \ast}({\hat k})\,
\end{eqnarray}
But we know that $\tilde u_{\nu \rightarrow}=\sum_{A} \tilde u_{\nu \rightarrow}^{(A)}$, so
\begin{eqnarray}
\nonumber
&&\sum_{r=1}^{3} [2k_{10}\,]^{-1/2}\,
\mathbf a_{\hat k, r} e_{\,r}^{\,\nu}({\hat k})=\\
&&=\sum_{A}\frac{1}{2}[2k_{1A}\,]^{-1/2}\,\left[
\mathbf c_{\hat k, A} \varepsilon_{\,A}^{\,\nu}({\hat k})\,(\frac{k_{1A}+k_{10}}{k_{10}}) - \mathbf c_{\hat k, A}^\dag \varepsilon_{\,A}^{\,\nu \ast}({\hat k})\,(\frac{k_{1A}-k_{10}}{k_{10}})\right]
\end{eqnarray}
Using the fact that $-\,g_{\,\mu\nu}\;e_{\,r}^{\,\mu}({\hat k})\,
e_{\,s}^{\,\nu}({\hat k})=\delta_{\,rs}$,
\begin{eqnarray}
\!\!\!\!\! \mathbf a_{\hat k, r}=\sum_{A}\frac{1}{2}\left[
-\mathbf c_{\hat k, A}\,g_{\,\mu\nu}\;e_{\,r}^{\,\mu}({\hat k})\, \varepsilon_{\,A}^{\,\nu}({\hat k})\,\frac{k_{1A}+k_{10}}{\sqrt{k_{10}k_{1A}}} + \mathbf c_{\hat k, A}^\dag \,g_{\,\mu\nu}\;e_{\,r}^{\,\mu}({\hat k})\,\varepsilon_{\,A}^{\,\nu \ast}({\hat k})\,\frac{k_{1A}-k_{10}}{\sqrt{k_{10}k_{1A}}}\right] .
\end{eqnarray}
One can see that the same equation we obtained in Sec. 3 (\ref{relation}). Thus, our results are consistent.

\subsection{Escaping \label{refl}}

Now we consider the case when the particle enters from the left half-space to the right one, so we have to take $\tilde u_{\mu \leftarrow}=0$.
\begin{eqnarray}
\tilde u_{\nu \rightarrow}^{(L)}=\tilde v_{\nu L \rightarrow} - \tilde v_{\nu L \leftarrow}\\
\tilde u_{\nu \rightarrow}^{(+)}=\frac{(\tilde v_{\nu + \rightarrow} - \tilde v_{\nu + \leftarrow})k_{1+}}{k_{10}}\\
\tilde u_{\nu \rightarrow}^{(-)}=\frac{(\tilde v_{\nu - \rightarrow} - \tilde v_{\nu - \leftarrow})k_{1-}}{k_{10}}
\end{eqnarray}
$\nu = 0, 2, 3$
\begin{eqnarray}
\tilde u_{\nu \rightarrow}^{(A)}=\tilde v_{\nu A \rightarrow} + \tilde v_{\nu A \leftarrow}
\end{eqnarray}
And now we can find which part is reflected:
\begin{eqnarray}
\tilde v_{\nu L \leftarrow}=0\\
\tilde v_{\nu + \leftarrow} = \frac{k_{1+}-k_{10}}{k_{1+}+k_{10}} \tilde v_{\nu + \rightarrow}\\
\tilde v_{\nu - \leftarrow} = \frac{k_{1-}-k_{10}}{k_{1-}+k_{10}} \tilde v_{\nu - \rightarrow}
\end{eqnarray}
and which part is passed through:
\begin{eqnarray}
\tilde u_{\nu \rightarrow}^{(L)}=\tilde v_{\nu L \rightarrow}\\
\tilde u_{\nu \rightarrow}^{(+)}=\frac{2k_{1+}}{k_{10}+k_{1+}}\tilde v_{\nu + \rightarrow}\\
\tilde u_{\nu \rightarrow}^{(-)}=\frac{2k_{1-}}{k_{10}+k_{1-}}\tilde v_{\nu - \rightarrow}
\end{eqnarray}
Finally, we have  the solution consisting of:\\
for $x_1>0$
\begin{eqnarray}
A_\mu=\int  \frac{\mathrm d^3 \hat k}{(2\pi)^3}\ \theta(\omega^2 - k^2_\bot - m^2) \!\ \tilde u_{\mu \rightarrow} (\omega,k_2,k_3) e^{ik_{10} x_1} e^{i \hat k \hat x}
\end{eqnarray}
and for $x_1<0$
\begin{eqnarray}
\left\lbrace
\begin{array}{cc}
A_1=\int  \frac{d^3 \hat k}{(2\pi)^3}\ \theta(\omega^2 - k^2_\bot - m^2) \!\ \tilde u_{1 \rightarrow}(\omega,k_2,k_3) e^{ik_{10} x_1} e^{i \hat k \hat x}
\\\\
A_\nu = \int  \frac{d^3 \hat k}{(2\pi)^3}\ \theta(\omega^2 - k^2_\bot - m^2) \!\ \sum_A \left[\tilde v_{\nu A \rightarrow}(\omega,k_2,k_3) e^{ik_{1A} x_1}+\tilde v_{\nu A \leftarrow}(\omega,k_2,k_3) e^{-ik_{1A} x_1}\right] e^{i \hat k \hat x}\,\,\,\,\,\,\nu=0,2,3\\
\end{array}\right.
\end{eqnarray}
There are only four independent coefficients responsible for the amplitude of the incoming particle.

We can also write our solution in such a form:
for $x_1<0$
\begin{eqnarray}
A_\nu = \int  \frac{\mathrm d^3 \hat k}{(2\pi)^3}\ \theta(\omega^2 - k^2_\bot - m^2) \!\ \sum_A \left[\tilde v_{\nu A \rightarrow}(\omega,k_2,k_3) e^{- i k x}+\tilde v_{\nu A \leftarrow}(-\omega,-k_2,-k_3) e^{i k x}\right] \,\,\,\,\,\,\,\,\,\,\,\,\,\, \\
\nonumber \nu=0,2,3
\end{eqnarray}

\subsection {Entrance}
It is also interesting to look at the case of entering the parity breaking medium. It means that our particle moves from the left half-space to the right one. Thus we can take $\tilde v_{\mu A \rightarrow}=0$. From eqs. (\ref{uv}) one derives,
\begin{eqnarray}
\tilde u_{\nu \rightarrow}^{(A)}=\frac{k_{10}-k_{1A}}{k_{10}+k_{1A}} \tilde u_{\nu \leftarrow}^{(A)} \label{uA}\\
\tilde v_{\nu A \leftarrow}=\frac{2k_{10}}{k_{10}+k_{1A}}\tilde u_{\nu \leftarrow}^{(A)}\label{vA}
\end{eqnarray}
The solutions are:
for $x_1>0$,
\begin{eqnarray}
\!\!\!\!\!\!\!\!\!\!\! A_\nu=\int  \frac{\mathrm d^3 \hat k}{(2\pi)^3}\ \theta(\omega^2 - k^2_\bot - m^2) \!\ \left[\tilde u_{\nu \leftarrow} (\omega,k_2,k_3) e^{-ik_{10} x_1}+ \tilde u_{\nu \rightarrow} (\omega,k_2,k_3) e^{ik_{10} x_1} \right]e^{i \hat k \hat x}
\end{eqnarray}
for $x_1<0$,
\begin{eqnarray}
A_\nu = \int  \frac{\mathrm d^3 \hat k}{(2\pi)^3}\ \theta(\omega^2 - k^2_\bot - m^2) \!\ \sum_A \tilde v_{\nu A \leftarrow}(\omega,k_2,k_3) e^{-ik_{1A} x_1} e^{i \hat k \hat x}\,\,\,\,\,\,\nu=0,2,3
\end{eqnarray}

The only point we don't know in this case is the relations between $\tilde u_{\nu \leftarrow}^{(+)},\tilde u_{\nu \leftarrow}^{(-)}, \tilde u_{\nu \leftarrow}^{(L)}$. We can find them using (\ref{v}), (\ref{uA}), (\ref{vA}) and all incoming amplitudes $(\tilde u_{\nu \leftarrow})$.
Below are presented the expressions for all components,
\begin{eqnarray}
\left\lbrace
\begin{array}{cc}
\tilde u_{0 \leftarrow}^{(L)}=\frac{\omega^2}{\omega^2-k_{\bot}^2}\tilde u_{0 \leftarrow}+\frac{\omega k_3}{\omega^2-k_{\bot}^2}\tilde u_{3 \leftarrow} + \frac{\omega k_2}{\omega^2-k_{\bot}^2}\tilde u_{2 \leftarrow}\\

\tilde u_{0 \leftarrow}^{(+)}=-\frac{k_{\bot}^2}{2(\omega^2-k_{\bot}^2)}\tilde u_{0 \leftarrow}-\frac{\omega k_3-i k_2 \sqrt{\omega^2 - k_{\bot}^2}}{2(\omega^2-k_{\bot}^2)}\tilde u_{3 \leftarrow}-\frac{\omega k_2+i k_3 \sqrt{\omega^2 - k_{\bot}^2}}{2(\omega^2-k_{\bot}^2)}\tilde u_{2 \leftarrow}\\

\tilde u_{0 \leftarrow}^{(-)}=-\frac{k_{\bot}^2}{2(\omega^2-k_{\bot}^2)}\tilde u_{0 \leftarrow}-\frac{\omega k_3+i k_2 \sqrt{\omega^2 - k_{\bot}^2}}{2(\omega^2-k_{\bot}^2)}\tilde u_{3 \leftarrow}-\frac{\omega k_2-i k_3 \sqrt{\omega^2 - k_{\bot}^2}}{2(\omega^2-k_{\bot}^2)}\tilde u_{2 \leftarrow}\\

\end{array}\right.
\end{eqnarray}

\begin{eqnarray}
\left\lbrace
\begin{array}{cc}
\tilde u_{2 \leftarrow}^{(L)}=-\frac{k_2^2}{\omega^2-k_{\bot}^2}\tilde u_{2 \leftarrow}-\frac{\omega k_2}{\omega^2-k_{\bot}^2}\tilde u_{0 \leftarrow} - \frac{k_2 k_3}{\omega^2-k_{\bot}^2}\tilde u_{3 \leftarrow}\\

\tilde u_{2 \leftarrow}^{(+)}=\frac{\omega^2 - k_3^2}{2(\omega^2-k_{\bot}^2)}\tilde u_{2 \leftarrow}+\frac{\omega k_2-i k_3 \sqrt{\omega^2 - k_{\bot}^2}}{2(\omega^2-k_{\bot}^2)}\tilde u_{0 \leftarrow}+\frac{k_2 k_3-i \omega \sqrt{\omega^2 - k_{\bot}^2}}{2(\omega^2-k_{\bot}^2)}\tilde u_{3 \leftarrow}\\

\tilde u_{2 \leftarrow}^{(-)}=\frac{\omega^2 - k_3^2}{2(\omega^2-k_{\bot}^2)}\tilde u_{2 \leftarrow}+\frac{\omega k_2+i k_3 \sqrt{\omega^2 - k_{\bot}^2}}{2(\omega^2-k_{\bot}^2)}\tilde u_{0 \leftarrow}+\frac{k_2 k_3+i \omega \sqrt{\omega^2 - k_{\bot}^2}}{2(\omega^2-k_{\bot}^2)}\tilde u_{3 \leftarrow}\\
\end{array}\right.
\end{eqnarray}

\begin{eqnarray}
\left\lbrace
\begin{array}{cc}
\tilde u_{3 \leftarrow}^{(L)}=-\frac{k_3^2}{\omega^2-k_{\bot}^2}\tilde u_{3 \leftarrow}-\frac{\omega k_3}{\omega^2-k_{\bot}^2}\tilde u_{0 \leftarrow} - \frac{k_2 k_3}{\omega^2-k_{\bot}^2}\tilde u_{2 \leftarrow}\\

\tilde u_{3 \leftarrow}^{(+)}=\frac{\omega^2 - k_2^2}{2(\omega^2-k_{\bot}^2)}\tilde u_{3 \leftarrow}+\frac{\omega k_3+i k_2 \sqrt{\omega^2 - k_{\bot}^2}}{2(\omega^2-k_{\bot}^2)}\tilde u_{0 \leftarrow}+\frac{k_2 k_3+i \omega \sqrt{\omega^2 - k_{\bot}^2}}{2(\omega^2-k_{\bot}^2)}\tilde u_{2 \leftarrow}\\

\tilde u_{3 \leftarrow}^{(-)}=\frac{\omega^2 - k_2^2}{2(\omega^2-k_{\bot}^2)}\tilde u_{3 \leftarrow}+\frac{\omega k_3-i k_2 \sqrt{\omega^2 - k_{\bot}^2}}{2(\omega^2-k_{\bot}^2)}\tilde u_{0 \leftarrow}+\frac{k_2 k_3-i \omega \sqrt{\omega^2 - k_{\bot}^2}}{2(\omega^2-k_{\bot}^2)}\tilde u_{2 \leftarrow}\\
\end{array}\right.
\end{eqnarray}
\subsection{Green's function \label{green} }
Let's build the Green's function (the propagator) for each polarization separately. We can say that our equations (\ref{systFourier}) have two linearly independent solutions, for example, one  for the particle moving from the left half-space, and another one  for the particle moving from the right half-space.
\begin{eqnarray}
\psi_1^{(A)}=\left[\frac{2k_{1A}}{k_{10}+k_{1A}}\theta(x_1)e^{ik_{10}x_1}+\theta(-x_1)(e^{ik_{1A}x_1}+\frac{k_{1A}-k_{10}}{k_{1A}+k_{10}}e^{-ik_{1A}x_1})\right]\tilde v_{\nu A \rightarrow}\\
\psi_2^{(A)}=\left[\frac{2k_{10}}{k_{10}+k_{1A}}\theta(-x_1)e^{-ik_{1A}x_1}+\theta(x_1)(e^{-ik_{10}x_1}-\frac{k_{1A}-k_{10}}{k_{1A}+k_{10}}e^{ik_{10}x_1})\right]\tilde u_{\nu \leftarrow}^{(A)}
\end{eqnarray}
Using the conventional Sturm-Liouville theory we can determine  the Green function after the Fourier transformation in $x_0,x_2,x_3$,
\begin{eqnarray}
G^{(A)}(\omega,k_2,k_3,x_1,x_1')=-\frac{1}{W(\psi_1^{(A)},\psi_2^{(A)})}
\left\lbrace
\begin{array}{cc}
\psi_2^{(A)}(x_1')\psi_1^{(A)}(x_1), x_1\leq x_1'\\
\psi_1^{(A)}(x_1')\psi_2^{(A)}(x_1), x_1>x_1'
\end{array} \right.
\end{eqnarray}
The Wronskian is given by,
\begin{eqnarray}
W(\psi_1^{(A)}(x_1'),\psi_2^{(A)}(x_1'))=(\frac{4ik_{10}k_{1A}}{k_{10}+k_{1A}})\tilde u_{\nu \leftarrow}^{(A)}\tilde v_{\nu A \rightarrow},
\end{eqnarray}

Next let's perform the Fourier transformation in $x_1, x_1'$.
\begin{eqnarray}
&&\tilde G^{(A)}(\omega,k_2,k_3,k_1,k_1')=\int\int_{-\infty}^{\infty}\mathrm d x_1 \mathrm d x'_1 i \frac{k_{10}+k_{1A}}{4k_{10}k_{1A}}\times
\nonumber\\
&&\times(\theta(x_1'-x_1)\psi_2^{(A)}(x_1')\psi_1^{(A)}(x_1)+\theta(x_1-x_1')\psi_1^{(A)}(x_1')\psi_2^{(A)}(x_1))e^{-ik_1x_1}e^{-ik_1'x_1'} =
\nonumber
\end{eqnarray}
\begin{eqnarray}
&&=(i\frac{k_{10}+k_{1A}}{4k_{10}k_{1A}})\{\frac{2k_{1A}}{k_{10}+k_{1A}}\left[\frac{1}{i}v.p.\frac{1}{k_1'+k_{10}}+\frac{1}{i}v.p.\frac{1}{k_1+k_{10}}\right]
\left[\frac{1}{i(k_1'+k_1-i\varepsilon)}\right]-
\nonumber \\
&&- \frac{2k_{1A}(k_{1A}-k_{10})}{(k_{10}+k_{1A})^2}\left[\frac{1}{i}v.p.\frac{1}{k_1'-k_{10}}+\frac{1}{i}v.p.\frac{1}{k_1-k_{10}}\right]
\left[\frac{1}{i(k_1'+k_1-2k_{10}-i\varepsilon)}\right]+
\nonumber \\
&&+\frac{2k_{10}}{k_{10}+k_{1A}}\left[\frac{1}{i}v.p.\frac{1}{k_1'+k_{1A}}+\frac{1}{i}v.p.\frac{1}{k_1+k_{1A}}\right]
\left[\frac{i}{k_1'+k_1+i\varepsilon}+\frac{k_{1A}-k_{10}}{k_{1A}+k_{10}}(\frac{i}{k_1'+k_1+2k_{1A}+i\varepsilon})\right]+
\nonumber\\
&&+\frac{2k_{1A}}{k_{10}+k_{1A}}\left[\pi\delta(k_1'+k_{10})-\frac{k_{1A}-k_{10}}{k_{1A}+k_{10}}\pi\delta(k_1'-k_{10}) \right]\left[\frac{1}{i(k_1-k_{10}-i\varepsilon)}\right]+
\nonumber\\
&&+\frac{2k_{1A}}{k_{10}+k_{1A}}\left[\pi\delta(k_1+k_{10})-\frac{k_{1A}-k_{10}}{k_{1A}+k_{10}}\pi\delta(k_1-k_{10}) \right]\left[\frac{1}{i(k_1'-k_{10}-i\varepsilon)}\right]+
\nonumber\\
&&+\left[\frac{1}{i(k_1'+k_{10}-i\varepsilon)}-\frac{k_{1A}-k_{10}}{k_{1A}+k_{10}}(\frac{1}{i(k_1'-k_{10}-i\varepsilon)})-\frac{2k_{10}}{k_{1A}+k_{10}}\frac{1}{i}v.p.\frac{1}{k_1'+k_{1A}}\right]\times
\nonumber\\
&&\times
\left[\frac{i}{k_1-k_{1A}+i\varepsilon}+\frac{k_{1A}-k_{10}}{k_{1A}+k_{10}}(\frac{i}{k_1+k_{1A}+i\varepsilon})\right]+
\nonumber\\
&&+\left[\frac{1}{i(k_1+k_{10}-i\varepsilon)}-\frac{k_{1A}-k_{10}}{k_{1A}+k_{10}}(\frac{1}{i(k_1-k_{10}-i\varepsilon)})-\frac{2k_{10}}{k_{1A}+k_{10}}\frac{1}{i}v.p.\frac{1}{k_1+k_{1A}}\right]\times
\nonumber\\
&&\times
\left[\frac{i}{k_1'-k_{1A}+i\varepsilon}+\frac{k_{1A}-k_{10}}{k_{1A}+k_{10}}(\frac{i}{k_1'+k_{1A}+i\varepsilon})\right]
\}
\end{eqnarray}

\section{Conclusions and outlook}

For massive photons and vector mesons the physical subspace consists of the three polarizations
$A=\pm, L$, but the chiral transverse states with complex polarization
vectors $\varepsilon^{\,\nu}_{\,\pm}(k_{\x\pm})$ propagate as massive states with effective masses dependent on the wave vector $k_{\x\pm}$.
If we take the photon mass $m_\gamma=0$
one of helicity states becomes superluminal for all values of the momenta and produces a
sort of Cherenkov radiation \cite{Lehnert:2004be}, gradually splitting into three  photons with negative
polarizations (see \cite{pick,klink} for similar arguments for space-like background
vectors). We have to stress that, kinematically, the high-energy photon with positive
polarization can {\em also} undergo splitting into the negative polarization photons.
Both splittings are kinematically allowed as it can be easily read out from the
inequality for the forward decay (we neglect here the photon mass), \be \omega_\pm
({\bf k})=\sqrt{{\bf k}^2\pm \zeta |{\bf k}|} > 3\,
\omega_-(\frac{{\bf k}}{3}) . \ee Thus if the phenomenon of positron (dilepton) excess is
accounted for by the instability of photons in a pseudoscalar background an
accompanying effect might be the suppression of high-energy $\gamma$ rays from the same
region, depending on the value of the effective photon mass, bearing in mind that this
process is anyway a one-loop effect.
In addition, there is the
possibility of ``radiative'' LIV decays $e^-\to e^- \gamma$; the momentum threshold
being $|k_1| >  4m^2_e / \zeta_x$. This effect will change the energy spectrum of
the $e^+ e^-$ pair produced in LIV $\gamma \to e^+ e^-$ decays, but it is suppressed by
a power of $\alpha$ and the cross-section must be proportional to $\zeta_x$ too \cite{axion2,Espriu:2010bj,Espriu:2010ip}. In spite of that axion density in galaxies is very low
there might be the regions of relatively dense axion localization ("axion stars" \cite{mielke}) and to detect them one has to know  how in-medium photons and leptons leave those regions. Thus boundary effects are crucial for their discovery.
The influence of a boundary between parity-odd medium and vacuum on the decay width of photons and vector mesons represents also a very interesting problem for calculation of realistic yield of dileptons produced by $\rho,\omega$ mesons in central heavy-ion collisions when local parity breaking occurs \cite{aaep}. Thus it certainly
deserves to be a subject of further investigation, in particular, of quantization for time-like CS vector, of transmission and reflection of unstable photons and vector bosons against a boundary and their relative dependence on energies, at last, of peculiarities of angular dependence allowing to detect their emission from axion clouds or fireballs.
\acknowledgments
This work is partially supported by Grants RFBR 09-02-00073-a and by SPbSU grant 11.0.64.2010. A.A.A. acknowledges also the financial support from projects FPA2010-20807, 2009SGR502, CPAN (Consolider CSD2007-00042). S.S.K.\ is supported by Dynasty Foundation
stipend. We are grateful to Yu.M.Pis'mak for valuable comments and to P. Giacconi for her assistance at early stage of our work.
\section*{Appendix A}
Let's integrate eq.\eqref{gensol} from $-\varepsilon$ to $\varepsilon$ taking into account following relations:
\begin{eqnarray}
\nonumber
&&\int_{-\varepsilon}^{\varepsilon} \mathrm d x_1 \tilde A_{\nu} =\int_{-\varepsilon}^{0} \mathrm d x_1 \tilde A_{\nu} +\int_{0}^{\varepsilon} \mathrm d x_1 \tilde A_{\nu} \\
\nonumber\\ \nonumber
&&\int_{-\varepsilon}^{\varepsilon} \mathrm d x_1  \theta(-x_1) \tilde A_{\nu} =\int_{-\varepsilon}^{0} \mathrm d x_1 \tilde A_{\nu} \\
\nonumber \\ \nonumber
&&\int_{0}^{\varepsilon} \mathrm d x_1 \tilde A_{\nu} =\frac{-1}{i k_{10}}(\tilde u_{\nu \rightarrow}-\tilde u_{\nu \leftarrow})+\frac{\tilde u_{\nu \rightarrow} e^{i k_{10} \varepsilon}}{i k_{10}}+\frac{\tilde u_{\nu \leftarrow} e^{-i k_{10} \varepsilon}}{-i k_{10}}\\
\nonumber \\ \nonumber
&&\int_{-\varepsilon}^{0} \mathrm d x_1  \tilde A_{\nu} = \sum_{A} [\frac{1}{i k_{10}}(\tilde v_{\nu A \rightarrow}-\tilde v_{\nu A \leftarrow})-\frac{\tilde v_{\nu A \rightarrow} e^{-i k_{10} \varepsilon}}{i k_{10}}-\frac{\tilde v_{\nu A \leftarrow} e^{i k_{10} \varepsilon}}{i k_{10}}]\\
\nonumber \\ \nonumber
&&\int_{-\varepsilon}^{\varepsilon} \mathrm d x_1  \partial_1^2 \tilde A_{\nu} =\partial_1 \tilde A_\nu |_{-\varepsilon}^{\varepsilon}=\nonumber \\ \nonumber
&&=i k_{10} \tilde u_{\nu \rightarrow} e^{i k_{10} \varepsilon}-i k_{10} \tilde u_{\nu \leftarrow} e^{-i k_{10} \varepsilon}-\sum_A [i k_{1A} \tilde v_{\nu A \rightarrow} e^{-i k_{1A} \varepsilon}-i k_{1A} \tilde v_{\nu A \leftarrow} e^{i k_{1A} \varepsilon}],
\end{eqnarray}
and consider separately terms containing $e^{i k_{1A} \varepsilon}$.

Take now the first equation from the system \eqref{systFourier} and prove the next equality:\\\\
$(-\omega^2+m^2+k_\bot^2)\frac{\tilde v_{0 A \leftarrow}}{i k_{1A}}e^{i k_{1A} \varepsilon}-i k_{1A}\tilde v_{0 A \leftarrow}e^{i k_{1A} \varepsilon}=i \zeta_x (k_2 \frac{\tilde v_{2 A \leftarrow}}{i k_{1A}}e^{i k_{1A} \varepsilon} - k_3 \frac{\tilde v_{3 A \leftarrow}}{i k_{1A}}e^{i k_{1A} \varepsilon})$ .\\\\
To prove it, we multiply this equality by $i k_{1A}$:\\\\
$(k_{1A}^2+(\omega^2 + m^2 + k_\bot^2))\tilde v_{0 A \leftarrow}= i \zeta_x (k_2 \tilde v_{2 A \leftarrow} - k_3 \tilde v_{3 A \leftarrow})$\\\\
Finally using expression for $k_{1A}$ we get:\\\\
$\epsilon_A \sqrt{\omega^2 - k_\bot^2} \tilde v_{0 A \leftarrow}=i (k_2 \tilde v_{2 A \leftarrow} - k_3 \tilde v_{3 A \leftarrow})$,\\\\
where
$\epsilon_A=\left\lbrace
\begin{array}{cc}
1, A=+\\
-1, A=-\\
0, A=L
\end{array}
\right.$\\
and it coincides with relations \eqref{v}. Similarly, one can check vanishing of terms with $e^{- i k_{1A} \varepsilon}$ and $e^{\pm i k_{10} \varepsilon}$.


\begin{thebibliography}{2011}
\bibitem{Shore:2004sh}
  G.~M.~Shore,
  Nucl.\ Phys.\  B {\bf 717}, 86 (2005)
  [arXiv:hep-th/0409125].

\bibitem{Jacobson:2005bg}
  T.~Jacobson, S.~Liberati and D.~Mattingly,
  Annals Phys.\  {\bf 321}, 150 (2006)
  [arXiv:astro-ph/0505267].

\bibitem{Gamboa:2005pd}
  J.~Gamboa, J.~Lopez-Sarrion and A.~P.~Polychronakos,
  Phys.\ Lett.\  B {\bf 634}, 471 (2006)
  [arXiv:hep-ph/0510113].


\bibitem{Nozari:2007zzb}
  K.~Nozari and D.~Sadatian,
  Electron.\ J.\ Theor.\ Phys.\  {\bf 4}, 87 (2007).

\bibitem{Kostelecky:2008ts}
  V.~A.~Kosteleck\'y and N.~Russell,
  arXiv:0801.0287 [hep-ph].

\bibitem{Bietenholz:2008ni}
  W.~Bietenholz,
  arXiv:0806.3713 [hep-ph].

\bibitem{Stecker:2009hj}
  F.~W.~Stecker and S.~T.~Scully,
  New J.\ Phys.\  {\bf 11}, 085003 (2009)
  [arXiv:0906.1735 [astro-ph.HE]].

\bibitem{Alfaro:2009xc}
  J.~Alfaro and P.~Gonzalez,
  arXiv:0909.3883 [hep-ph].

\bibitem{Lehnert:2009pf}
  R.~Lehnert,
  J.\ Phys.\ Conf.\ Ser.\  {\bf 171}, 012036 (2009)
  [arXiv:0907.1319 [hep-ph]].

\bibitem{Shao:2010wk}
  L.~Shao and B.~Q.~Ma,
  Mod.\ Phys.\ Lett.\  A {\bf 25}, 3251 (2010)
  [arXiv:1007.2269 [hep-ph]].
\bibitem{Carroll:1989vb}
  S.~M.~Carroll, G.~B.~Field and R.~Jackiw,
  Phys.\ Rev.\  D {\bf 41}, 1231 (1990).

\bibitem{Jackiw:1999qq}
  R.~Jackiw,
  Int.\ J.\ Mod.\ Phys.\  B {\bf 14}, 2011 (2000)
  [arXiv:hep-th/9903044].

\bibitem{Volovik:1999up}
  G.~E.~Volovik,
  JETP Lett.\  {\bf 70}, 1 (1999)
  [Pisma Zh.\ Eksp.\ Teor.\ Fiz.\  {\bf 70}, 3 (1999)]
  [arXiv:hep-th/9905008].

\bibitem{AGS2002}
A.A. Andrianov, P. Giacconi and R. Soldati,
JHEP {\bf 02}, 030 (2002).
%

\bibitem{Klinkhamer:2004hg}
  F.~R.~Klinkhamer and G.~E.~Volovik,
  Int.\ J.\ Mod.\ Phys.\  A {\bf 20}, 2795 (2005)
  [arXiv:hep-th/0403037].

\bibitem{Ebert:2004pq}
  D.~Ebert, V.~C.~Zhukovsky and A.~S.~Razumovsky,
  Phys.\ Rev.\  D {\bf 70}, 025003 (2004)
  [arXiv:hep-th/0401241].

\bibitem{Zhukovsky:2005iu}
  V.~C.~Zhukovsky, A.~E.~Lobanov and E.~M.~Murchikova,
  Phys.\ Rev.\  D {\bf 73}, 065016 (2006)
  [arXiv:hep-ph/0510391].


\bibitem{Gamboa:2005bf}
  J.~Gamboa and J.~Lopez-Sarrion,
  Phys.\ Rev.\  D {\bf 71}, 067702 (2005)
  [arXiv:hep-th/0501034].


\bibitem{Das:2005uq}
  A.~K.~Das, J.~Gamboa, J.~Lopez-Sarrion and F.~A.~Schaposnik,
  Phys.\ Rev.\  D {\bf 72}, 107702 (2005)
  [arXiv:hep-th/0510002].


\bibitem{Itin:2007cv}
  Y.~Itin,
  Phys.\ Rev.\  D {\bf 76}, 087505 (2007)
  [arXiv:0709.1637 [hep-th]].


\bibitem{Gomes:2007rv}
  M.~Gomes, J.~R.~Nascimento, E.~Passos, A.~Y.~Petrov and A.~J.~da Silva,
  Phys.\ Rev.\  D {\bf 76}, 047701 (2007)
  [arXiv:0704.1104 [hep-th]].

\bibitem{Brito:2008ec}
  F.~A.~Brito, L.~S.~Grigorio, M.~S.~Guimaraes, E.~Passos and C.~Wotzasek,
  Phys.\ Rev.\  D {\bf 78}, 125023 (2008)
  [arXiv:0810.3180 [hep-th]].


\bibitem{Scarpelli:2008fw}
  A.~P.~B.~Scarpelli, M.~Sampaio, M.~C.~Nemes and B.~Hiller,
  Eur.\ Phys.\ J.\  C {\bf 56}, 571 (2008)
  [arXiv:0804.3537 [hep-th]].


\bibitem{Casana:2008nw}
  R.~Casana, M.~M.~.~Ferreira and C.~E.~H.~Santos,
  Phys.\ Rev.\  D {\bf 78}, 025030 (2008)
  [arXiv:0804.0431 [hep-th]].

\bibitem{Andrianov:1994qv}
  A.~A.~Andrianov and R.~Soldati,
  Phys.\ Rev.\  D {\bf 51}, 5961 (1995)
  [arXiv:hep-th/9405147].


\bibitem{Colladay:1996iz}
  D.~Colladay and V.~A.~Kosteleck\'y,
  Phys.\ Rev.\  D {\bf 55}, 6760 (1997)
  [arXiv:hep-ph/9703464].

\bibitem{Colladay:1998fq}
  D.~Colladay and V.~A.~Kosteleck\'y,
  Phys.\ Rev.\  D {\bf 58}, 116002 (1998)
  [arXiv:hep-ph/9809521].

\bibitem{Andrianov:1998wj}
  A.~A.~Andrianov and R.~Soldati,
  Phys.\ Lett.\  B {\bf 435}, 449 (1998)
  [arXiv:hep-ph/9804448].



\bibitem{Coleman:1997xq}
  S.~R.~Coleman and S.~L.~Glashow,
  Phys.\ Lett.\  B {\bf 405}, 249 (1997)
  [arXiv:hep-ph/9703240].

\bibitem{Coleman:1998en}
  S.~R.~Coleman and S.~L.~Glashow,
  arXiv:hep-ph/9808446.

\bibitem{Coleman:1998ti}
  S.~R.~Coleman and S.~L.~Glashow,
  Phys.\ Rev.\  D {\bf 59}, 116008 (1999)
  [arXiv:hep-ph/9812418].



\bibitem{Myers:2003fd}
  R.~C.~Myers and M.~Pospelov,
  Phys.\ Rev.\ Lett.\  {\bf 90} (2003) 211601
  [arXiv:hep-ph/0301124].

\bibitem{Montemayor:2005ka}
  R.~Montemayor and L.~F.~Urrutia,
  Phys.\ Rev.\  D {\bf 72}, 045018 (2005)
  [arXiv:hep-ph/0505135].

\bibitem{Kharlanov:2009pv}
  O.~G.~Kharlanov and V.~C.~Zhukovsky,
  Phys.\ Rev.\  D {\bf 81}, 025015 (2010)
  [arXiv:0905.3680 [hep-th]].



\bibitem{Kostelecky:2002hh}
  V.~A.~Kosteleck\'y and M.~Mewes,
  Phys.\ Rev.\  D {\bf 66}, 056005 (2002)
  [arXiv:hep-ph/0205211];\,
  Phys.\ Rev.\  D {\bf 80}, 015020 (2009)
  [arXiv:0905.0031 [hep-ph]].


\bibitem{Shore:2003zc}
  G.~M.~Shore,
  Contemp.\ Phys.\  {\bf 44}, 503 (2003)
  [arXiv:gr-qc/0304059].

\bibitem{Cheng:2006us}
  H.~C.~Cheng, M.~A.~Luty, S.~Mukohyama and J.~Thaler,
  JHEP {\bf 0605}, 076 (2006)
  [arXiv:hep-th/0603010].

\bibitem{Andrianov:1998ay}
  A.~A.~Andrianov, R.~Soldati and L.~Sorbo,
  Phys.\ Rev.\  D {\bf 59}, 025002 (1999)
  [arXiv:hep-th/9806220].

\bibitem{ArkaniHamed:2003uy}
  N.~Arkani-Hamed, H.~C.~Cheng, M.~A.~Luty and S.~Mukohyama,
  JHEP {\bf 0405}, 074 (2004)
  [arXiv:hep-th/0312099].

\bibitem{Jenkins:2003hw}
  A.~Jenkins,
  Phys.\ Rev.\  D {\bf 69}, 105007 (2004)
  [arXiv:hep-th/0311127].


\bibitem{Peloso:2004ut}
  M.~Peloso and L.~Sorbo,
  Phys.\ Lett.\  B {\bf 593}, 25 (2004)
  [arXiv:hep-th/0404005].


\bibitem{ArkaniHamed:2004ar}
  N.~Arkani-Hamed, H.~C.~Cheng, M.~Luty and J.~Thaler,
  JHEP {\bf 0507}, 029 (2005)
  [arXiv:hep-ph/0407034].

\bibitem{Alfaro:2009iv}
  J.~Alfaro and L.~F.~Urrutia,
  Phys.\ Rev.\  D {\bf 81}, 025007 (2010)
  [arXiv:0912.3053 [hep-ph]].


\bibitem{sikivie} L.Abbott and P. Sikivie, Phys. Lett. 120B, 133 (1983).

\bibitem{raffelt}M. Kuster, G. Raffelt and B. Beltran (eds), Axions: Theory, Cosmology
and Experimental Searches, Lecture Notes
in Physics 741 (2008).

\bibitem{wmap}E. W. Kolb and M. S. Turner, The Early Universe (Westview Press, 1990).

 \bibitem{wmap1} Y. Sofue and V. Rubin,
Ann. Rev. Astron. Astrophys. 39, 137 (2001).

\bibitem{wmap2} S. J. Asztalos et al., Ap. Jour. 571, L27 (2002).

\bibitem{mielke} F.E. Schunck and E.W. Mielke, Class.Quant.Grav.20:R301-R356,2003 ;\\
E. W. Mielke and J. A. Velez Perez, Phys.Lett.B671:174-178,2009.

\bibitem{axion1} A. A. Andrianov, D. Espriu, P. Giacconi and R. Soldati,
  JHEP {\bf 0909}, 057 (2009).

\bibitem{axion2}\ A. A. Andrianov, D. Espriu, F. Mescia and A. Renau,
  Phys.  Lett.  B {\bf 684}, 101 (2010) .

\bibitem{Espriu:2011vj}
  D.~Espriu and A.~Renau,
  arXiv:1106.1662 [hep-ph].

\bibitem{Espriu:2010bj}
  D.~Espriu and A.~Renau,
  arXiv:1010.3580 [hep-ph].

\bibitem{Espriu:2010ip}
  D.~Espriu, F.~Mescia and A.~Renau,
  JCAP {\bf 1108}, 002 (2011)
  [arXiv:1010.2589 [hep-ph]].

\bibitem{ceres} M. Masera, (HELIOS/3 Collaboration), Nucl.
Phys. A {\bf 590}, 103c (1995).

\bibitem{ceres1} G. Agakichiev et al.(CERES Collaboration),
Eur. Phys. J. C{\bf 4}, 231 (1998).

\bibitem{NA60} R. Arnaldi et al.[NA60 Collaboration], Phys. Rev. Lett.
{\bf 96}, 162302 (2006).

\bibitem{phenix} PHENIX Collaboration (A. Adare et al.), Phys. Rev. C{\bf 81}, 034911 (2010).

\bibitem{HADES} G.~Agakichiev et al. (HADES Collab.), Phys. Rev. Lett.  {\bf 98}, 052302 (2007) ;\ Phys.Lett. {\bf  B 663} (2008) 43.

\bibitem{lapidus} K. O. Lapidus and V. M. Emel'yanov, Phys. Part. Nucl. {\bf 40}, 29 (2009).

\bibitem{tserruya} I.Tserruya,
0903.0415 [nucl-ex].

\bibitem{aaep} A.~A.~Andrianov, V.~A.~Andrianov, D.~Espriu and X.~Planells,
"Abnormal dilepton yield from local parity breaking in heavy-ion
collisions," arXiv:1010.4688 [hep-ph];\, PoS, QFTHEP2010, 053 (2010) .

\bibitem{anesp} A.~A.~Andrianov and D.~Espriu,
  Phys. Lett.  B {\bf 663}, 450 (2008).

\bibitem{anesp1} A.~A.~Andrianov, V.~A.~Andrianov and D.~Espriu,
  Phys. Lett. B {\bf 678}, 416 (2009).

\bibitem{kharzeev} D. Kharzeev, R. D. Pisarski and M. H. G. Tytgat,  Phys. Rev. Lett. {\bf 81}, 512 (1998).

\bibitem{zhit} K. Buckley, T. Fugleberg, and A. Zhitnitsky, Phys. Rev. Lett. {\bf 84}, 4814 (2000).

\bibitem{kharzeev1} D. Kharzeev, Phys.Lett. B{\bf 633}, 260 (2006).

\bibitem{kharzeev2}
D. E. Kharzeev, L. D. McLerran and H. J. Warringa, Nucl. Phys. A{\bf  803}, 227 (2008) .

\bibitem{lattice}
P. Buividovich, M. Chernodub, E. Luschevskaya and M. Polikarpov, Phys. Rev.  D {\bf 80}, 054503 (2009) .

\bibitem{star} B.~I.~Abelev {\it et al.}  [STAR Collaboration],
  Phys. Rev. Lett.  {\bf 103}, 251601 (2009).

\bibitem{star2}  S. A. Voloshin,
  J. Phys. Conf. Ser.  {\bf 230}, 012021 (2010) .

%
\bibitem{AACGS2010}
J. Alfaro, A.A. Andrianov, M. Cambiaso, P. Giacconi  and R. Soldati,
Int. J. Mod. Phys. A {\bf 25}, 3271 (2010);\,
Phys. Lett. B{\bf 639}, 586 (2006).
%
\bibitem{AEGS2009}
A.A. Andrianov, D. Espriu, P. Giacconi and R. Soldati,
JHEP {\bf 0909}, 057 (2009).
%

\bibitem{KL2001}
V.A.~Kosteleck\'y and R.~Lehnert,
Phys. Rev. D  {\bf 63}, 065008 (2001).

\bibitem{Lehnert:2004be}
  R.~Lehnert and R.~Potting,
  Phys.\ Rev.\  D {\bf 70}, 125010 (2004)
  [Erratum-ibid.\  D {\bf 70}, 129906 (2004)]
  [arXiv:hep-ph/0408285].

\bibitem{pick} V. Alan Kosteleck\'y and Austin G. M. Pickering,  Phys. Rev. Lett. 91, 031801 (2003).

\bibitem{klink}C. Adam and F.R. Klinkhamer, Nucl. Phys. B 657, 214 (2003).
\end{thebibliography}
\end{document}